\documentclass{pasj01}
\usepackage[T1]{fontenc}
\UseRawInputEncoding
\usepackage{amsmath}

\usepackage{comment}
\usepackage{mathrsfs}
\usepackage{booktabs}

\usepackage{graphicx}
\usepackage{support-caption}
\usepackage{subcaption}

\Received{$\langle$2021/12/15$\rangle$}
\Accepted{$\langle$2022/03/28$\rangle$}
\Published{$\langle$publication date$\rangle$}
\begin{document}

\title{ 
%\LETTERLABEL %%% <-- uncomment for LETTER article  
%\REVIEWLABEL %%% <-- uncomment for REVIEW article  
X-ray Selected Narrow-Line Active Galactic Nuclei in the COSMOS Field: Nature of Optically Dull Active Galactic Nuclei }

%%% begin:list of authors
% Do NOT capitalize all letters in "textsc".
\author{Itsna K \textsc{Fitriana}\altaffilmark{1}}
%\thanks{Example: Present Address is xxxxxxxxxx}}
\altaffiltext{1}{Astronomical Institute, Tohoku University, 6-3, Aramaki, Aoba-ku, Sendai, Miyagi, 980-8578, Japan}
\email{itsnafitriana@astr.tohoku.ac.jp}

\author{Takashi \textsc{Murayama},\altaffilmark{1}}
%\altaffiltext{}{B-Address of Institute}
%\email{bbbbb@xxx.xxx.xx.xx}

%\author{C-Firstname \textsc{C-Familyname}\altaffilmark{3}}
%\altaffiltext{3}{C-Address of Institute}
%\email{ccccc@xxx.xxx.xx.xx}
%%% end:list of authors

%% `\KeyWords{}' always has to be placed before ``\maketitle'' 
%%  List of Key Words:  https://academic.oup.com/pasj/pages/Pasj_Keywords 

\KeyWords{galaxies:active --- galaxies:nuclei --- galaxies: evolution --- galaxies: star formation}

\maketitle

\begin{abstract}
X-ray emission detection in a galaxy is one of the efficient tools for selecting Active Galactic Nuclei (AGNs). However, many X-ray-selected AGNs are not easily selected as AGNs by their optical emission. These galaxies, so-called optically dull (OD) AGNs, are fascinating since their X-ray emission is bright even though the AGN signature in the optical regime is absent. In a deep multiwavelength survey over 2 deg$^2$ of Cosmic Evolution Survey (COSMOS) field, we have looked for the OD AGNs using photometric, spectroscopic, and X-ray data. We identified 310 non-broad line sources with optical spectra as AGN using X-ray selection up to redshift $z\sim 1.5$. We inspected the spectra to check for any AGN signature in their optical emission lines: [Ne\,\emissiontype{V}] forbidden emission line, Mass Excitation diagram (MEx), color excitation diagram (TBT), and excess in [O\,\emissiontype{II}] emission line. Finally, we found 48 AGNs show AGN signatures in optical spectrum classified as narrow-line AGN  and 180 AGNs that did not show any AGN signature as OD AGN sample. The simple explanation of OD AGN's nature is due to a bright host galaxy that dilutes the AGN light or due to dust materials obscuring the AGN light. We found that the bright host galaxy dilution explains nearly $70\%$ of our OD AGN sample. At the same time, the dust material obscuration is unlikely for the main reason. By estimating the Eddington ratio, we also found that 95/180 of our OD AGNs have a lower accretion rate of  $(\lambda_\text{Edd})\lesssim 10^{-2}$ than the typical AGN value. We expected the lower accretion rate sources that suffer from neither host galaxy dilution nor obscuration to have Radiatively Inefficient Flow (RIAF) in their accretion disk. Finally, nine sources have been identified to be most likely host the RIAF disk.
\end{abstract}

%\pagebreak
%\linenumbers

\section{Introduction}
\label{sec:Intro}
% ***************************************************
% Introduction
% ***************************************************

Most galaxies, particularly massive galaxies with stellar mass of $M_* \sim 10^{10}$ $M_{\odot}$ -- $10^{12}$ $M_{\odot}$, host a supermassive black hole (SMBH) at their nucleus \citep{Kormendy__1995}. Essentially all SMBHs in massive galaxies are thought to experience actively growth of their SMBH, during which they are observed as active galactic nuclei (AGN) \citep{Marconi2004}. The AGN is now believed to be 
powered supply by the accretion matter to the SMBH emits comparable or even more radiation than the rest of the host galaxy.

The optical spectra observation generally classifies AGNs as type 1, also known as broad line AGN (BL AGN), and type 2 or narrow line AGN (NL AGN). BL AGN shows broad and narrow emission lines on their spectra, while NL AGN only presents narrow ones. The lack of a broad emission line of NL AGN is attributed to a dusty structure obscuring the central engine and the broad-line region (BLR). The straightforward interpretation supports the idea that BL AGN and NL AGN are built intrinsically by the same object but observed in a different line of sight. This model is well known as the traditional unification model introduced by \citet{Antonucci1993}.

Unfortunately, a simple unified model has a limit explaining the whole observations of AGN since it is based solely on geometric obscuration. By definition, the NL AGN will show a hidden BLR through a spectropolarimetry observation. Otherwise, many NL AGNs show no broad emission line even in very deep spectropolarimetric observation. Deep X-ray surveys also have revealed ``optically dull'' (OD) AGNs (\cite{Comastri_2002}; \cite{Merloni_2013}; \cite{Trump2009a}), which emit bright X-ray emission but lack of optical AGN signature. The nature of OD AGNs could be explained by dilution from the prominent host galaxies' light (\cite{Moran_2002}; \cite{Trump2009a}; \cite{Pons2016}) or obscuration by dusty material near the central engine or beyond the host galaxies(\cite{Civano_2007}; \cite{Rigby_2006}). The exciting nature is an explanation that the OD AGNs could have a different intrinsic structure that caused a lower accretion rate than typical AGN (\cite{Yuan_2004}; \cite{Trump_2011b}).

An AGN selection through only one wavelength observation could result in a more biased sample than a multi-wavelength observation. It is due to the AGN accretion process that produces an extensive range of electromagnetic emissions. Principally, X-ray selection of AGN will be highly efficient compared to optical selection. It is due to the X-ray selection less suffer from obscuration of the intervening material and dilution of light from the host galaxy than in optical wavelength. So therefore, the interesting question about these OD AGNs is; How can the nuclei produce high X-ray emission while the optical evidence of AGN is absent? Could we expect to a particular stage of AGN evolutionary process?

Observations suggest a lower limit in accretion rates of $L/L_\text{Edd}\ge 0.01$ for BLR formation in AGNs (\cite{Trump_2011b}). By the absence of BLR formation, it becomes clearer to explain the optical dullness among the OD AGNs regarding the absence of AGN signature when we observe through optical spectrum observation.  \citet{Narayan_1998} introduced A model of radiatively inefficient accretion flow (RIAF) (\cite{Narayan_1998}; \cite{Narayan_2012}) to explain the lower accretion rate disk nature. They suggest that the normal thin disk (\cite{Shakura1973}) will be truncating into a geometrically thick and optically thin RIAF disk at smaller radii near the SMBH. Such object is predicted to have weak UV/optical emission and lack strong emission lines. This property is usually observed among the optically weak low-luminosity AGNs (\cite{Nemmen_2006}; \cite{Narayan_2012}) and some found in X-ray bright OD AGNs (\cite{Yuan_2004}; \cite{Trump_2011b}).

The OD AGN may represent a significant AGN population. Moreover, the mismatch between X-ray and optical observation may also question whether orientation can be the only parameter to distinguish BL AGN and NL AGN \citep{Padovani_2017}. Therefore, the study of OD AGN is essential for building a complete census of AGN over a broad range in luminosity which will give a better understanding of the AGN physical structure and the accretion physics.

This paper based mainly on the X-ray and spectroscopic data of the COSMOS survey. We take advantage of the large potential of the COSMOS database that provide deep observational data. We could analyze fainter sources compare to the previous study (i.e. \cite{Rigby_2006},\cite{Trump_2011},\cite{Pons2016}) thanks to the work of the COSMOS team. Increasing of OD AGN number is crucial for understanding their nature and their role in the galaxy evolution.

This paper presents the sample of X-ray AGNs in the Cosmic Evolution Survey (COSMOS) field with further spectral type analysis as NL AGN and OD AGN. Section \ref{sec:XrayAGN} describes the data sources and the selection of our OD AGN and NL AGN samples. Here, we explain our sample selection tools, the AGN signature that rises in the X-ray and optical spectrum. In Section \ref{sec:Results} we present the difference of our NL AGN and OD AGN sample in the host galaxy properties and X-ray properties. Section \ref{sec:natureODAGN} discusses the possible reasons for the optical dullness of our sample. Finally, we summarize our work in Section \ref{sec:summary}. We adopt a cosmology with $h=0.70$, $\Omega_M=0.3$, and $\Omega_{\Lambda}=0.7$ throughout.

\section{Data and AGN Selection}
\label{sec:XrayAGN}
%%%%%%%%%%%%%%%%%%%%%%%%%%%%%%%%%%%%%%%

\begin{table*}
    %\small
    \tbl{Catalog in COSMOS project that used in this work.}{%
    \begin{tabular}{l|llll}
        \hline
        \hline
        \textbf{Catalog  Name}  & \textbf{Wavelength} &$I_{AB}$ &  \textbf{Data used} & \textbf{Reference}\\
         \hline
         C-COSMOS legacy  & X-ray (0.5-10 keV) & $\le$24.6 &  X-ray properties &\citet{Marchesi2016}\\
         COSMOS2015  & Multi-band &$\le$25-26 & Photometry,&  \citet{Laigle2016}\\
          & observation& & SED fitting products & \\
         Z-COSMOS  & $5500-9659$ \AA &$\le$22.5-25 & Optical spectrum & \citet{Lilly_2007}\\
         Deimos10K  & $5500-9800$ \AA &$\le$23-25 & Optical spectrum & \citet{Hasinger_2018}\\
         ZEST Morphology  & $\sim4000-6700$ \AA & $\le$24 & Morphological & \citet{Scarlata_2007}\\
          & & &  poperties &  \\
         %non spectra X-ray AGN  & spectra data not available & 537 &  43.42 $\pm$ 0.57\\
        \hline
    \end{tabular}}\label{table:catalog}% is used to refer this table in the tex
\end{table*} 

\begin{figure*}
    \centering
	\includegraphics[width=1.4\columnwidth]{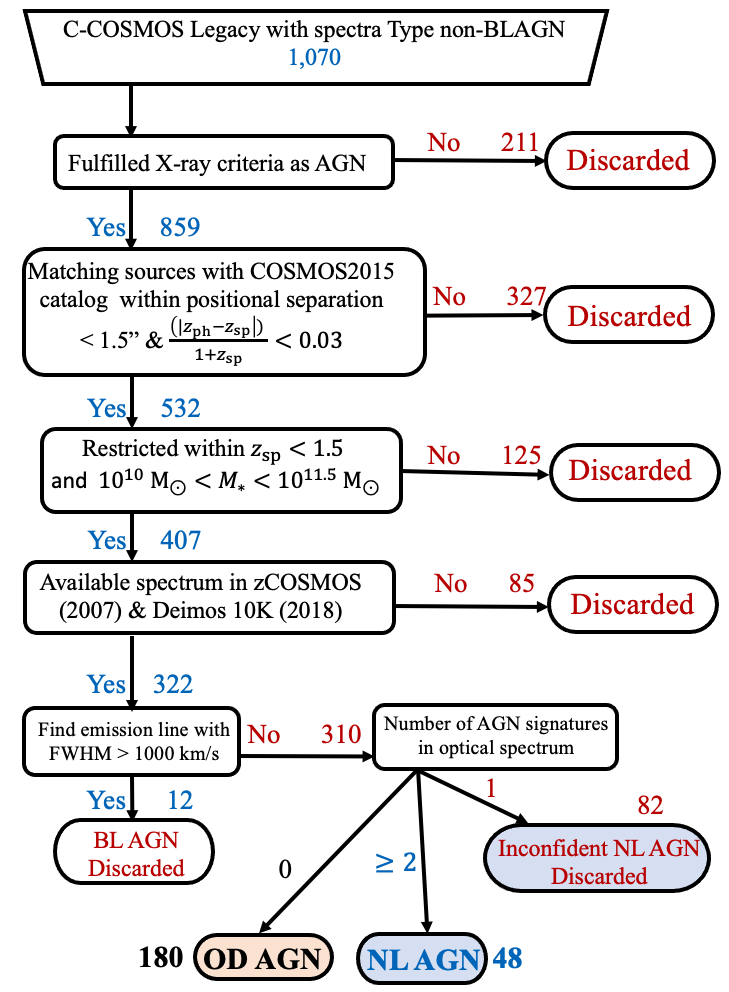}
    \caption{A flow chart of our sample selection scheme. See text for details}
    \label{fig:flowchart}
\end{figure*}

Multiwavelength data is necessary to understand OD AGNs better. For that purpose, we used a set of photometric and spectroscopic data from the Cosmic Evolution Survey (COSMOS) project. COSMOS is the extensive multiwavelength ground and space-based observations of the area 2 deg$^{2}$ of the sky \citep{Scoville_2007}. We adopted the X-ray data from Chandra COSMOS (C-COSMOS) legacy catalog \citep{Marchesi2016}, the photometry and spectral energy distribution (SED) products in UV, optical, and near-infrared wavelength from the COSMOS2015 catalog \citep{Laigle2016}, the optical spectrum data from zCOSMOS and DEIMOS10K catalog, and the optical morphological catalog from the Zurich Estimator of Structural Types (ZEST) catalog  \citep{Scarlata_2007}. We summarize the fourth COSMOS catalogs used in this work in Table \ref{table:catalog}. Meanwhile, the whole process of our AGN data selection is summarized in a flowchart Fig. \ref{fig:flowchart}.

\subsection{X-ray selection of AGN without broad line signature}
\label{sec:XAGNselection}

The C-COSMOS catalog by \citet{Marchesi2016} contains 4016 X-ray sources down to flux limits of $2.2 \times 10^{-16}$ erg cm$^{-2}$ s$^{-1}$, $1.5 \times 10^{-16}$ erg cm$^{-2}$ s$^{-1}$, and $8.9 \times 10^{-16}$ erg cm$^{-2}$ s$^{-1}$ in energy bands of $0.5-2$ keV, $2-10$ keV, and $0.5-10$ keV, respectively. We used the X-ray luminosity, hardness ratio ($HR$), hydrogen column density ($N_\text{H}$) included in the catalog. Among them, 1770 sources are available with reliable spectroscopic redshifts ($z_\text{sp}$) and spectral type information within the COSMOS collaboration. With spectral type information of 632 sources are classified as BL AGN as showing at least one broad emission line ($FWHM$ $>$ 2000 km s$^{-1}$) in their optical spectra. We eliminated these BL AGNs from our sample. Other 1070 sources are classified as non-broad-line AGNs (non-BL AGNs) with only a narrow emission line or no emission line. We should note that the ``non-BL AGN'' class includes ``non-AGN'' galaxies (i.e., star-forming (SF) galaxies). It is due to the C-COSMOS catalog did not further separate them from NL AGNs. 
Instead of emission line classification of optical spectra, we adopted criteria following \citet{Trump2009a} to eliminate SF galaxies,
\begin{equation}
    L_{0.5-10 \text{keV}} > 3 \times 10^{42} \text{erg s}^{-1},
	\label{eq1}
\end{equation}

or 

\begin{equation}
     -1 \leq X/O \leq 1,
	\label{eq2}
\end{equation}

where \begin{gather*} X/O = \log f_X/f_O = \log f_{0.5-2 \text{keV}} +i_\text{AB}/2.5 + 5.352, \end{gather*} 
where $f_{0.5-2 \text{keV}}$ is in the unit of erg cm$^{-2}$ s$^{-1}$ while $i_\text{AB}$ is in the unit of mag. The luminosity limit in equation \ref{eq1} sets these constraints on X-ray luminosity in local SF galaxies. Meanwhile, equation \ref{eq2} is known as traditional ``X-ray AGN locus" of ~\cite{MAccacaro1998} which show typical X-ray to optical ratio value of AGN. These criteria are reliable in selecting AGN without contamination of powerful X-ray sources in SF galaxies (i.e., X-ray binary systems and ultra-luminous X-ray sources)typically with $L_X \sim 10^{39}-10^{41}$ erg s$^{-1}$. The sources were defined as X-ray AGN if they satisfy at least one of the two criteria above. Finally, we found 859 galaxies satisfied as X-ray AGNs. 

\begin{figure}
	\includegraphics[width=1\columnwidth]{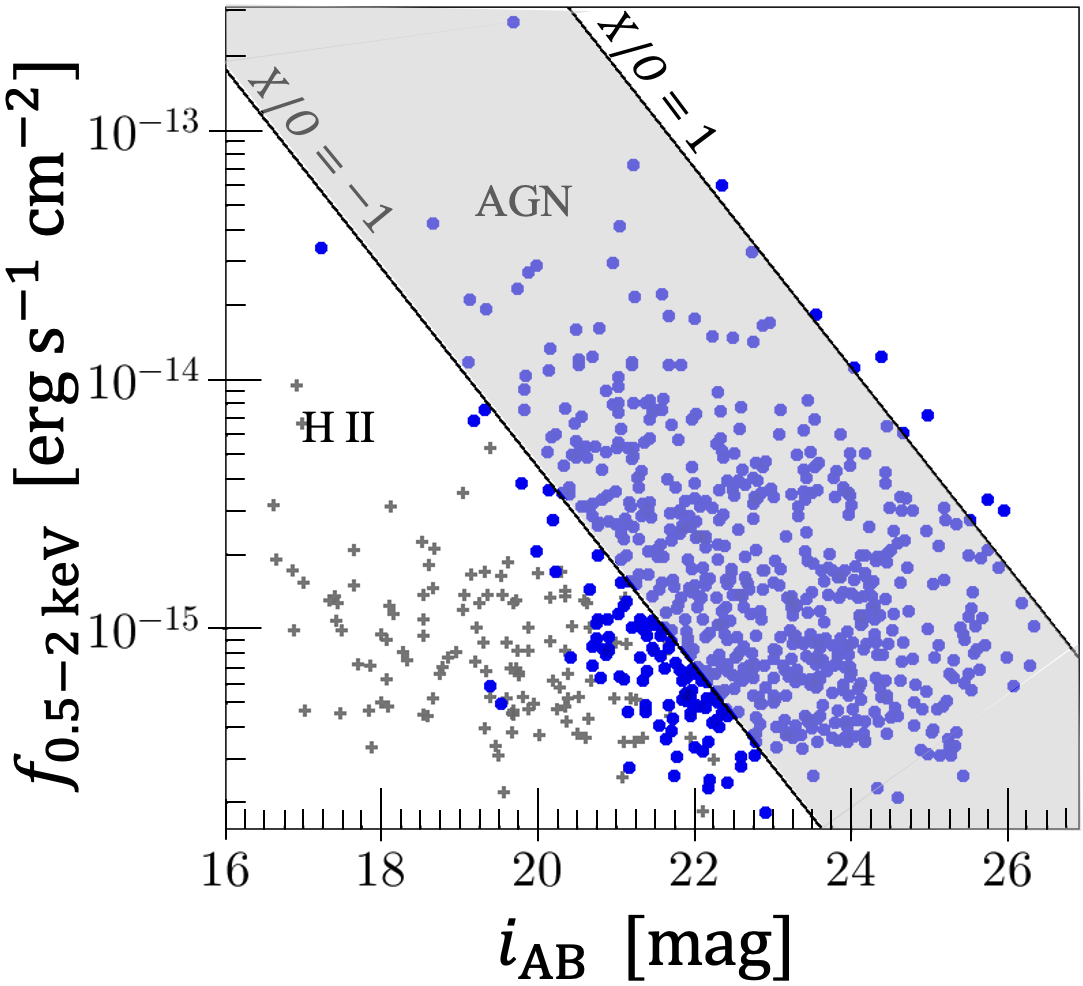}
	\centering
    \caption{Soft X-ray flux vs. i-band magnitude, for non-BL AGN sample. The grey shaded region defines the AGN locus with an optical X-ray ratio (X/O) between -1 and 1. Blue dots are the selected X-ray AGN, and the grey cross symbol is the remaining source. Only three sources have lower X-ray luminosity but are located in the AGN locus.}
    \label{fig:xolocus}
\end{figure}

In Fig. \ref{fig:xolocus}, blue dots indicate X-ray AGNs that satisfy at least one criterion of \ref{eq1} or \ref{eq2}. The AGN locus shows the criterion  \ref{eq2} in the grey shaded area between the two black solid lines ($-1\le X/O \le1$). Both criteria show mutually supportive results as many as 567 sources (66$\%$). Meanwhile, only 2$\%$ of the sources only meet the second criterion. Having two criteria to select the AGN will allow more samples to be studied.

\subsection{Redshift and stellar mass restriction}
\label{sec:redmass}

After selecting 859 X-ray AGNs, we matched them with sources in the COSMOS 2015 catalog. The catalog presents a newer version of the precise photometric redshift ($z_\text{ph}$) and contains more than half-million sources in the field of the COSMOS survey. The catalog also derived the absolute magnitudes, stellar mass ($M_*$), star formation rates ($SFR$), and other valuable properties of host galaxies. They did SED fitting of the 30-band photometry data with the galaxy SED template. These photometric data samples were obtained from hundreds of hours of telescope time in different surveys: UV survey of \textit{GALEX}(\cite{Zamojski2007}), optical 20 bands using Supreme-Cam instrument of Subaru (\cite{Taniguchi2007};\cite{Taniguchi2015}), Near-Infrared survey with the VIRCAM instrument as part of the UltraVISTA survey (\cite{Mccracken2012}), and Mid-IR by \textit{Spitzer} cycle 2 COSMOS survey (\cite{Sanders2007})). Those are crucial ingredients to discover the precise $z_\text{ph}$.

Matching process of our sample with COSMOS2015 catalog was performed both with (1) coincidence of the central positions in the C-COSMOS catalog and the COSMOS 2015 catalog within 1.5$^{\prime\prime}$, and (2) coincidence of the spectral redshifts ($z_\text{sp}$) in the C-COSMOS catalog and the photometric redshifts ($z_\text{ph}$) in the COSMOS 2015 catalog within $|z_\text{sp}-z_\text{ph}|/(1+z_\text{sp})<0.03$. Resulting number of the sample is 532.

Next, we restricted our sample in the redshift up to ($z< 1.5$) and stellar mass ($10^{10}$ $M_{\odot}$ $<$ $M_*<10^{11.5}$ $M_{\odot}$). We want to draw an evenly distributed sample and reject the bias analysis due to the observation limit. The redshift range is also required for finding the emission line needed in the optical AGN classification. These redshift and stellar mass limits reduced the number of samples to 407.

\subsection{Optical spectrum selection of narrow line AGN}
\label{sec:Opticalclass}

We used galaxy optical spectra available from the z-COSMOS catalog \citep{Lilly_2007} and DEIMOS 10K spectroscopic survey catalog \citep{Hasinger_2018}. Both catalogs were being undertaken in the COSMOS field, giving us a good cross-matched with the C-COSMOS legacy catalog. Z-COSMOS is a large-redshift survey that uses 600 hours of observation with the VIMOS spectrograph on the 8 m VLT. In the z-COSMOS catalog, we found the spectrum of 149 galaxies among the 407 X-ray AGNs remaining in the previous selection. Meanwhile, \citet{Hasinger_2018} presented the first comprehensive spectroscopic observations with the Deep Imaging Multi-Object Spectrograph (DEIMOS) on the Keck II telescope in the wavelength range of $\sim 5500 - 9800$ \AA. Of our X-ray AGNs, the optical spectrums of 197 galaxies were found in the Deimos 10K catalog. In total, we found 322 spectrum galaxies from both spectroscopy catalogs, with 24 sources overlapping in both catalogs.

We further inspected the available spectra of 322 galaxies to check the existence of AGN signatures in them. At first, we looked whether broad emission lines are still present and the galaxies which presented the emission lines broader than 1000 km s$^{-1}$. This threshold rejects all bonafide broad-line objects and reduces conventional narrow-line Seyfert 1 (NLS1) galaxies contamination  \citep{Caccianiga_2007}. The NLS1, which is part of BL AGN, can have line widths narrower than FWHM$<2000$ km s$^{-1}$. We found 12 galaxies with broad lines larger than 1000 km s$^{-1}$ in their spectra and discarded them from our sample. Therefore, they were not classified as broad line AGN in the C-COSMOS catalog.

The most commonly used diagnostic tools to separate AGN and SF population by optical emission is the BPT diagram \citep{BPT81}, which use the [O\,\emissiontype{III}]$\lambda$5007/H\emissiontype{$\beta$} and [NII]$\lambda$6584/H\emissiontype{$\alpha$} line ratios. Since the redshift of our sample is mostly $z\gtrsim 0.3$ , popular emission lines used in the BPT analysis (i.e. [N\,\emissiontype{II}]$\lambda$6584 and H\emissiontype{$\alpha$}) were shifted out of the optical spectral range. Therefore, alternative indicators have been proposed to find the AGN signature in optical spectrum. Particularly, we searched for the [Ne\,\emissiontype{V}] forbidden emission line at $\lambda =3426$ \AA { }, which is a good indicator of AGN activity (\cite{Heckman_2014}; \cite{Pouliasis_2020}; \cite{Padovani_2017}). We also adopted the Mass Excitation diagram (MEx) that uses the [O\,\emissiontype{III}]$\lambda$5007/H\emissiontype{$\beta$} versus $M_*$ (\cite{Juneau2011}; \cite{Juneau_2014}), the color excitation diagram (TBT) that use the [Ne\,\emissiontype{III}]$\lambda$3869/[O\,\emissiontype{II}]$\lambda$3727 versus $(g-z)$ rest-frame color \citep{Trouille_2011}. In addition, we used the novel method by \citet{Tanaka_2012}, that uses [O\,\emissiontype{II}] luminosity to separate AGN contribution from star-forming galaxies. 

We have applied these four diagnostic tools based on the emission line and galaxy properties to look for any galaxy with no signature in all methods. We marked them as OD AGN. Meanwhile, the sources classified as AGN by at least two diagnostic tools classified as NL AGN. Unconfident NL AGNs showed an AGN signature by only one diagnostic tool. We did not include them as NL AGN due to a lack of confidence. For our optical spectrum analysis, we used the latest version of the \texttt{SPECUTILS} \citep{astropy:2018} package in \texttt{PYTHON}. For the emission line used in the diagnostic tools, we measured emission lines in the optical spectra considering a single gaussian model. The threshold value of line detection is a signal-to-noise ratio of $S/N>3$ to get a trustworthy sample classification. 

\subsubsection{The [Ne\,\emissiontype{V}] emitters}
\label{sec:NeVemit}

The [Ne\,\emissiontype{V}]$\lambda3426$ has the exact physical origin as the [O\,\emissiontype{III}]$\lambda5007$, arising from photoionized gas in the narrow-line region (NLR). This line arises well outside the region of the heaviest obscuration (\cite{Padovani_2017}). The ionizing potential of  [Ne\,\emissiontype{V}] is 97 eV, which may only come from high energy sources as the NLR of AGN. \citet{Mignoli_2013} identified NL AGN that emits [Ne\,\emissiontype{V}], then compared them to the AGN which selected by X-ray selection and those from blue line ratio diagnostics. They concluded that the AGN that emits [Ne\,\emissiontype{V}] were able to identify low-luminosity and heavily obscured AGNs. 

Among our X-ray AGN sample, there are 47 X-ray AGNs that clearly show [Ne\,\emissiontype{V}] emission lines in their spectrum. 
\subsubsection{The MEx diagram}
\label{sec:MEx diagram}

The Mass-Excitation (MEx) diagnostic tool has been proposed by \citet{Juneau2011}. Instead of using the [N\,\emissiontype{II}]$\lambda$6584/H\emissiontype{$\alpha$} line ratio as in the classic BPT diagram, the MEx diagram uses the galaxy's stellar mass as a surrogate. The MEx technique successfully distinguishes the emission line originating from SF or AGN, possibly dealing with composite galaxies, which lie in the part of the diagram between the AGN and SF galaxies.

We used the revised two demarcation line as written in \citet{Juneau_2014} (see equation (1) and (2) in their paper). Among 26 sources with reliable measurements of [O\,\emissiontype{III}]$\lambda$5007/H\emissiontype{$\beta$} and stellar mass, we found that 24 sources are classified as AGN by the MEx diagram method. The remaining two sources are located precisely in the SFG region, which gives us no sources placed as composite galaxies.

\subsubsection{The TBT Diagram}
\label{sec:TBT diagram}

\citet{Trouille_2011} has been constructed the diagram with sample using in the Sloan Digital Sky Survey (SDSS) that is based on the $g-z$ rest-frame optical color ($^0(g-z)$) as a function of the ratio of  [Ne\,\emissiontype{III}]$\lambda3869$ and [O\,\emissiontype{II}]$\lambda3727$. The optical classification can be extended to a redshift of $z \sim$1.4  using only these two emission lines. Moreover, this diagram has the advantage of being little affected by reddening because the [Ne\,\emissiontype{III}]$\lambda3869$ and [O\,\emissiontype{II}]$\lambda3727$ are relatively close in wavelength. There is only one criterion for selecting AGN as follows:

\begin{equation}
    ^{0}(g-z)>-1.2 \times \log \text{([Ne\,\emissiontype{III}]$\lambda3869$/[O\,\emissiontype{II}]$\lambda3727$)}-0.4.
	\label{eq3}
\end{equation}

Our sample shows that 54 sources fall inside the AGN area by this criterion. In comparison, only four sources lie in the star-forming region. This method can be applied to a significant fraction of our samples due to the emission line availability in a wide redshift range. Besides, it is pretty easy to use, and this method gives significant overlap with the other diagnostic tools.

\subsubsection{The High [O\,\emissiontype{II}] Luminosity}
\label{sec:hiOII }

This technique was developed by \citet{Tanaka_2012} to study their low-luminosity AGN sample in the SDSS archive data. It can be applied at high redshift without making prior assumptions about host galaxy properties. \citet{Tanaka_2012} demonstrated a comparison between observed luminosity with that attributed to star formation to distinguish whether a galaxy host an AGN or not. 

The original “Oxygen-excess method” uses the total emission line that comes from [O\,\emissiontype{II}]$\lambda3727$ and [O\,\emissiontype{III}]$\lambda5007$. Even so, we use only the [O\,\emissiontype{II}]$\lambda3727$ line regarding the distant objects in our sample mostly showed only the [O\,\emissiontype{II}]$\lambda3727$ emission line. 

The [O\,\emissiontype{II}]$\lambda3727$ emission can arise from the star-forming region (SF region), and the NLR in a galaxy hosts an AGN. In other words, the observed emission is a combination of SF and AGN, or we can write as,
\begin{equation}
     L_{\text{[O\,\emissiontype{II}],obs}}=L_{\text{[O\,\emissiontype{II}],SF}}+L_{\text{[O\,\emissiontype{II}],AGN}}
\end{equation}
where $L_{\text{[O\,\emissiontype{II}],obs}}$ is the [O\,\emissiontype{II}] luminosity measured directly from the observation, $L_{\text{[O\,\emissiontype{II}],SF}}$ and $L$[O\,\emissiontype{II}]$_\text{AGN}$ are the contribution of [O\,\emissiontype{II}] emission coming from the host galaxy and AGN respectively. 

\begin{figure}
	\includegraphics[width=\columnwidth]{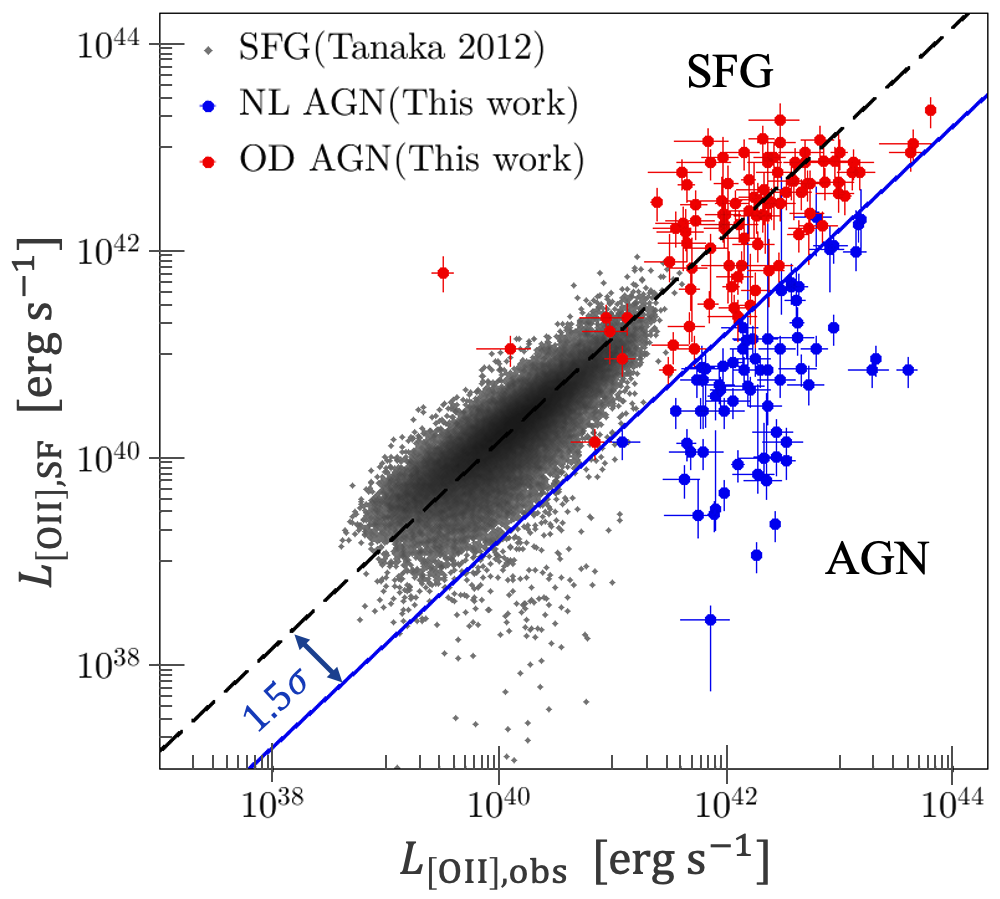}
    \caption{The relation between [O\,\emissiontype{II}]$\lambda3727$ luminosity, which measured directly from the observed spectrum, and the one that derived from the star formation rate ($SFR$). The region above the solid blue line shows the typical emission line arising from the SF region. In contrast, the region under the blue line shows the excess of [O\,\emissiontype{II}]$\lambda3727$emission line in the galaxy, probably coming from the AGN at the most. The grey dots are the distribution of Star-Forming Galaxies (SFG) in \citet{Tanaka_2012} work.}
    \label{fig:hiOII}
\end{figure}

Fig. \ref{fig:hiOII} shows the comparison between the [O\,\emissiontype{II}] luminosity that is directly measured from the optical spectrum ( $L_{\text{[O\,\emissiontype{II}],obs}}$) and the [O\,\emissiontype{II}] luminosity estimated from the $SFR$ ($L_{\text{[O\,\emissiontype{II}],SF}}$). Here, \citet{Tanaka_2012} show a systematic offset between the observed luminosity and the SF estimation luminosity for SF galaxies which is shown in a black dashed line as $\log L_{\text{[O\,\emissiontype{II}],obs}}=\log L_{\text{[O\,\emissiontype{II}],SF}}+0.16$ with a scatter of $\sigma=0.23$. Then, the original definition of the AGN selection by this method is shown by the excess of $L_{\text{[O\,\emissiontype{II}],obs}}$ to the $L_{\text{[O\,\emissiontype{II}],SF}}$ by $ > 1.5\sigma$ (shown by the blue solid line) after considering the offset. Thus, we can define the AGN sample as those with,

\begin{equation}
    \label{eq:hiOII_eq}
    \log\frac{L_{\text{[O\,\emissiontype{II}],obs}}}{L_{\text{[O\,\emissiontype{II}],SF}}}-0.16 >1.5\sigma,%\\
    %\log\frac{L_{\text{[O\,\emissiontype{II}]},obs}}{L_{\text{[O\,\emissiontype{II}]},SF}}> 0.16 + 1.5 \times 0.23 =0.505,
\end{equation}

The $L_{\text{[O\,\emissiontype{II}],obs}}$ in this work were directly measured from the available spectrum, which had been calibrated to the photometry flux. Meanwhile, the $L_{\text{[O\,\emissiontype{II}],SF}}$ is calculated from SFR equation following \citet{Kennicutt_1998} as described below,

\begin{equation}
    L{\text{[O\,\emissiontype{II}]}} ( 10^{41}\text{erg \ s}^{-1})=\frac{SFR (\text{M}_\odot\text{ yr}^{-1})}{(1.4 \pm 0.4)}
\end{equation}

After all, we found 68 sources laid below the solid blue line in Fig. \ref{fig:hiOII}, and thus they show an AGN excess in [O\,\emissiontype{II}] emission line. Meanwhile, 97 sources do not show the AGN excess in [O\,\emissiontype{II}] emission line, which typically comes from only their host galaxies.

The correction of dust attenuation is required for measuring the [O\,\emissiontype{II}] luminosity from the spectroscopic observation. The Balmer decreement is commonly used to estimate and correct the dust attenuation of emission line. However, there are only small number of sample with H\,\emissiontype{$\alpha$} and H\,\emissiontype{$\beta$} measurement by optical spectra due to redshift of our sample are higher than 0.3. Instead, we used dust extinction by stellar SED fitting provided in the COSMOS2015 catalog and estimated extinction for ionized gas region. Since the [O\,\emissiontype{II}]$\lambda 3727$ place in the $U$-band range, amount of dust extinction for [O\,\emissiontype{II}] is the same as that for U-band ($A_{[O\,\emissiontype{II}]}\sim A_U$). We applied a relation from \citet{kwangho2015} to obtain the extinction of gas in U as $A_\text{[O\,\emissiontype{II}]} \sim A_{U,gas} = 0.91 +(1.64 /0.44) (B-V)_*$, where $(B-V)_*$ is the color of stellar continuum. Finally we derived that $A_\text{[O\,\emissiontype{II}]}$ is between  $\sim 0.9-1.75$, thus dereddened [O\,\emissiontype{II}] luminosity would be 2-5 times higher. the dust absorption could reduce observed [O\,\emissiontype{II}] luminosity to $\sim 0.1-0.4$ times of the unreddened emission. The correction will be diverse for each sources. However dust correction would only slightly change the number of NL AGNs. Moreover, applying dust extinction correction for ionized gas from stellar SED fitting has large ambiguity, and thus we decided not to apply this dust correction for [O\,\emissiontype{II}] luminosity.
%

%Three sources show a sizeable upper error in $L_{\text{[O\,\emissiontype{II}]}, SF}$. However, we counted them still as an AGN since the main target of this research is the OD AGN population. The OD AGN sample will become a more clean sample by this definition. 

\subsection {Final sample of OD AGNs and NL AGNs}
\label{sec:finsam}

\begin{table*}
    \tbl{The terms used to define various type of AGN in this work}{
    \begin{tabular}{l   l l l l l l}
        \hline
        \hline
        \textbf{Sample}  & \textbf{Optical spectra} & \textbf{Total} & \textbf{$z_\textbf{sp}$} &\textbf{$\log M_*$} & \textbf{$\log SFR$} & \textbf{$\log L_{0.5-10\text{keV}}$}\\
        \hline
         OD AGN & no AGN signatures  & 180 & $0.34-1.488$ & $10.78 \pm 0.3$& $1.01 \pm 0.96$& $43.18 \pm 0.41$\\
         %missclassified OD AGN & no AGN signatures  & 180 & $0.34-1.488$ & $10.78 \pm 0.3$& $1.01 \pm 0.96$& $43.18 \pm 0.41$\\
         NL AGN  & $\ge 2$ AGN signatures& 48& $0.511-1.478$ & $10.82 \pm 0.32$& $0.48 \pm 1.1$&  43.37 $\pm$ 0.42\\
         unconfident NL AGN$^*$  & 1 AGN signature & 82 & $0.317-1.449$ & $10.78 \pm 0.33$& $0.75 \pm 1.12$&  43.28 $\pm$ 0.48\\
         BL AGN$^*$  & FWHM $\ge 1000$ km s$^{-1}$ & 12 & $0.1333-1.3226$ & $10.44 \pm 0.25$& $1.69 \pm 1.16$&  43.66 $\pm$ 0.61\\
        \hline
         \multicolumn{7}{l}{$^*$ The sample will not be used for the discussion in section 4 and follows.}\\
         %non spectra X-ray AGN  & spectra data not available & 537 &  43.42 $\pm$ 0.57\\
    \hline
    \end{tabular}}
    \label{table:AGNterm} % is used to refer this table in the text 
\end{table*} 

\begin{figure}
\centering
	\includegraphics[width=\columnwidth]{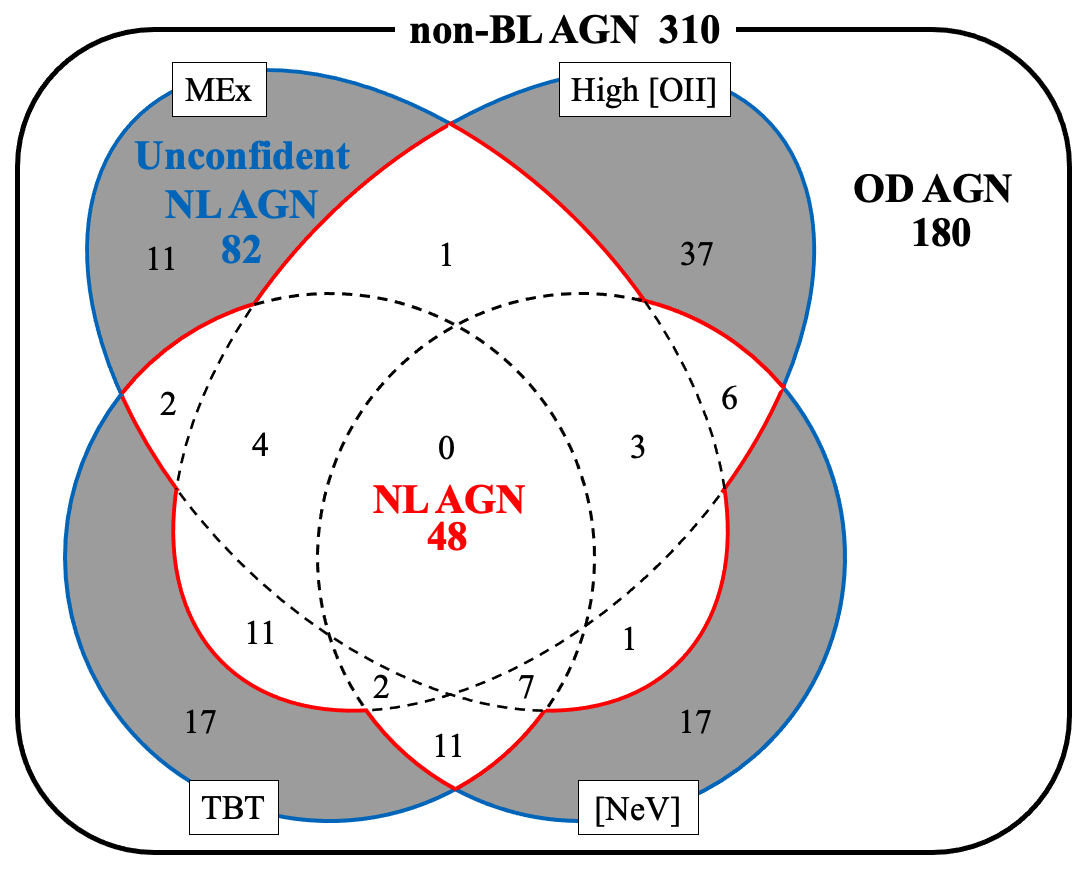}
    \caption {The result of the optical classification method in Section \ref{sec:Opticalclass} is drawn in the above Venn diagram. The NL AGNs are described inside the light region, which shows that sources are selected as AGN by more than one diagnostic tool. Otherwise, the shaded region shows the unconfident NL AGNs selected as AGN by only one diagnostic tool. The number of OD AGNs can be seen in the upper right, which consists of sources that can not show AGN signatures in their optical spectrum. The number of 310 shows the total spectroscopical data used in this study, eliminating all the broad emission line sources. }
    \label{fig:venn}
\end{figure}

Our optical spectrum selection for NL AGNs in section \ref{sec:Opticalclass} is shown by Venn diagram as in Fig. \ref{fig:venn}. A source is defined as NL AGN if it is classified as AGN by more than one diagnostic tool(shown in the white region in Fig. \ref{fig:venn}). Then, 48 (15\%) sources were classified as NL AGN. The term NL AGN in the subsequent analysis refers to this sample definition.

On the other hand, we found 180 sources to be OD AGN sample. The OD AGN sample is combined from 101 sources that misclassified as SF galaxies and also from 79 sources that were not able to be drawn in the above diagnostic tool. We found that majority of the misclassified OD AGNs were coming from the [O\,\emissiontype{II}] emission classification, which is 97 sources. Meanwhile, the undetected OD AGN is defined as a source that does not show an acceptable emission line (signal-to noise-ratio below the threshold) to be drawn in the optical diagnostic tools. Since all of our sample are ranges between redshift of 0.3 $< z_\text{sp} <$ 1.5, at least one of the emission lines used under this work ([Ne\,\emissiontype{V}]$\lambda3426$, [O\,\emissiontype{II}]$\lambda3727$, [Ne\,\emissiontype{III}]$\lambda3869$, [H\,\emissiontype{$\beta$}]$\lambda4861$, and [O\,\emissiontype{III}]$\lambda5007$) should be within the spectroscopic observations. After all these tests to find evidence for AGN in optical spectra, the OD AGN is defined as a source that do not show any AGN signatures in optical regime.

Meanwhile, we also found sources selected as AGN by only one diagnostic tool classified as unconfident sources (illustrated in the greyed area). We found 82 (15\%) sources to be unconfident NL AGNs. These unconfident NL AGNs will be discarded in the following work.  

Table \ref{table:AGNterm} show the redshift range of the final sample. The NL AGN are distributed in the redshift range of $0.511 \le z_\text{sp}\le 1.478$ while the OD AGNs ranges in the $0.34 \le z_\text{sp}\le 1.488$. Our X-ray AGN yields in the higher redshift even for OD AGN, which was not showing any AGN signature in optical, is most likely due to the spectroscopic data availability. There is also not much difference in the stellar mass distribution, which gives an unbias sample regarding the galaxy's properties. However, we could still find that most OD AGNs give a higher $SFR$ and lower X-ray luminosity than NL AGN. 

\section[Results]{Results}
\label{sec:Results}
As illustrated in Fig. \ref{fig:flowchart}, we finally obtained 48 NL AGNs and 180 OD AGNs samples. Due to the redshift range in our sample, [O\,\emissiontype{II}] emission line is the most observed in our sources. Therefore, we compare the [O\,\emissiontype{II}] luminosity for the NL AGN and OD AGN samples in this section. There are 47 NL AGNs as [O\,\emissiontype{II}] emitters and only one NL AGN that does not emit [O\,\emissiontype{II}] line. Meanwhile, there are 97 OD AGNs that show [O\,\emissiontype{II}] emission line. 

Besides the emission line comparison, we also compare host galaxy properties between NL AGN and OD AGN samples. We use the axial ratios of the host galaxies from the COSMOS morphological catalog, which is obtained from the Zurich Estimator of Structural Types (ZEST; \citet{Scarlata_2007}) catalog. 

\begin{figure*}
\begin{subfigure}{.5\textwidth}
  \centering
  \caption{}
  % include first image
  \includegraphics[width=\textwidth]{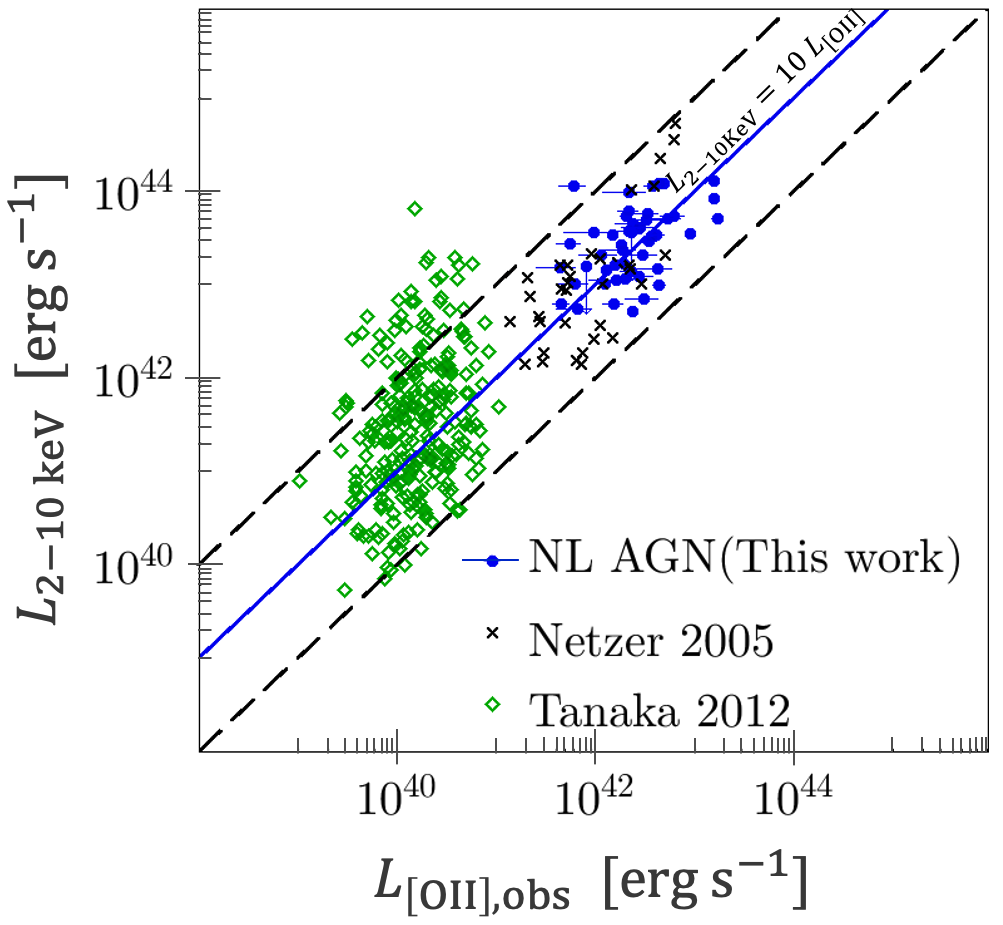}  
  \label{fig:5a}
\end{subfigure}
\begin{subfigure}{.37\textwidth}
  \centering
  \caption{}
  % include second image
  \includegraphics[height=1.2\textwidth]{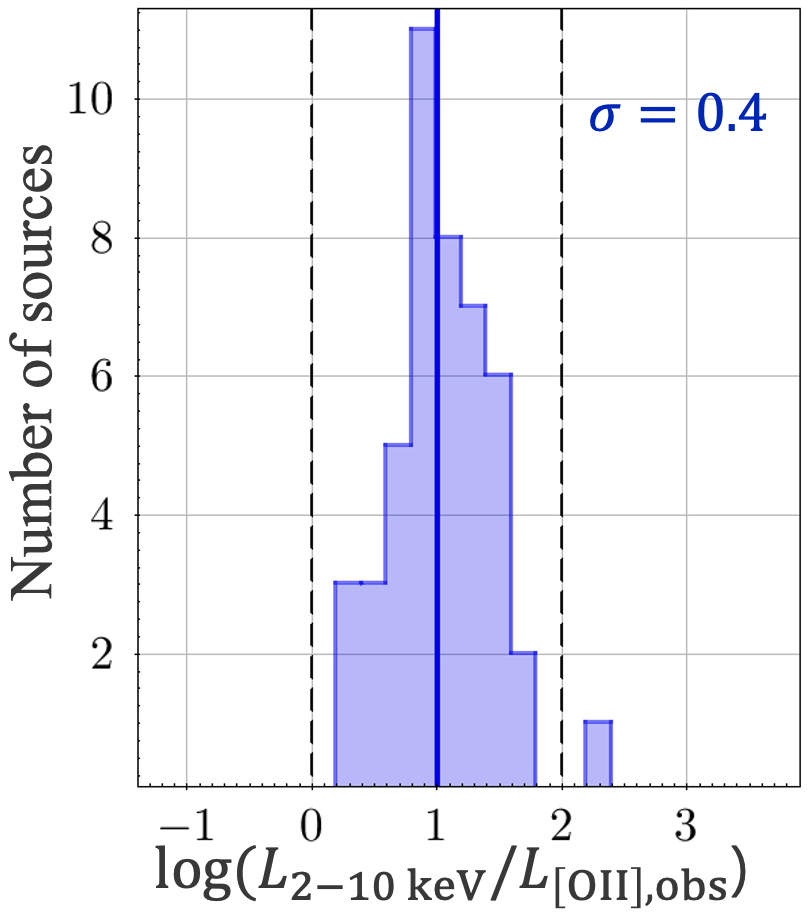}  
  \label{fig:5b}
\end{subfigure}

\begin{subfigure}{.5\textwidth}
  \centering
  \caption{}
  % include third image
  \includegraphics[width=\textwidth]{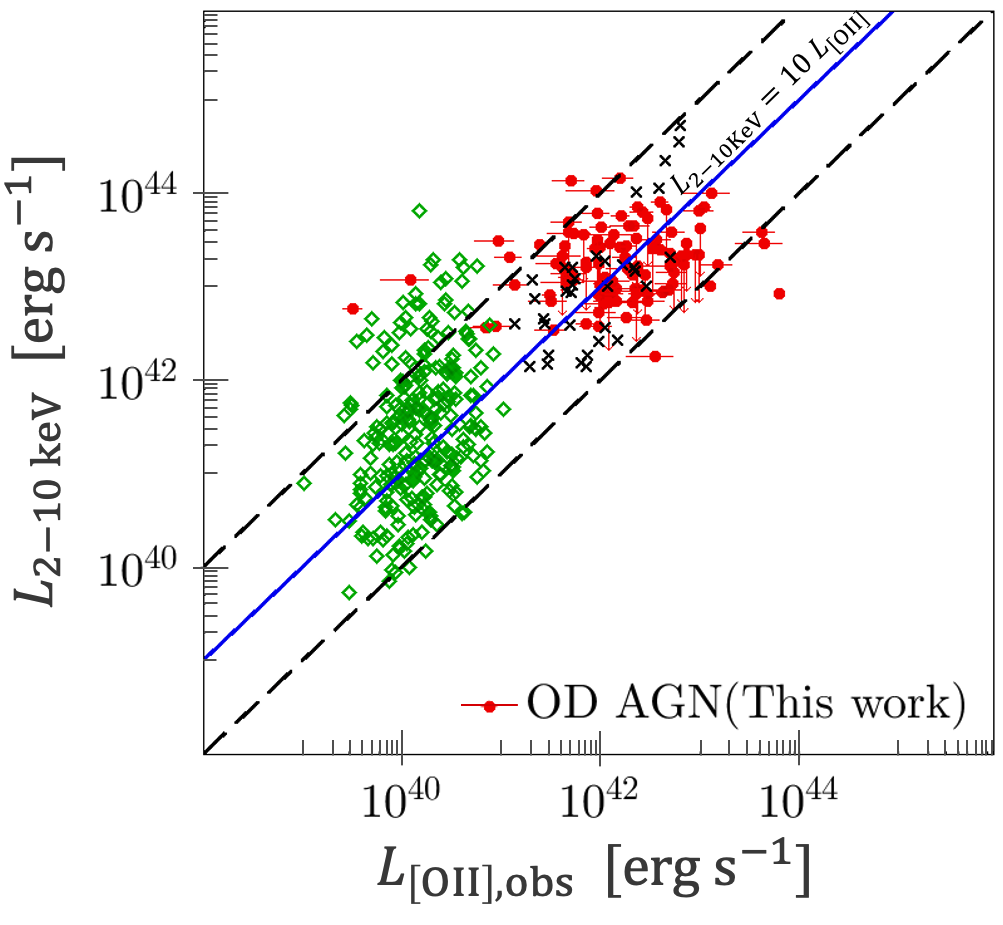}  
  
  \label{fig:5c}
\end{subfigure}
\begin{subfigure}{.37\textwidth}
  \centering
  \caption{}
  % include fourth image
  \includegraphics[height=1.2\textwidth]{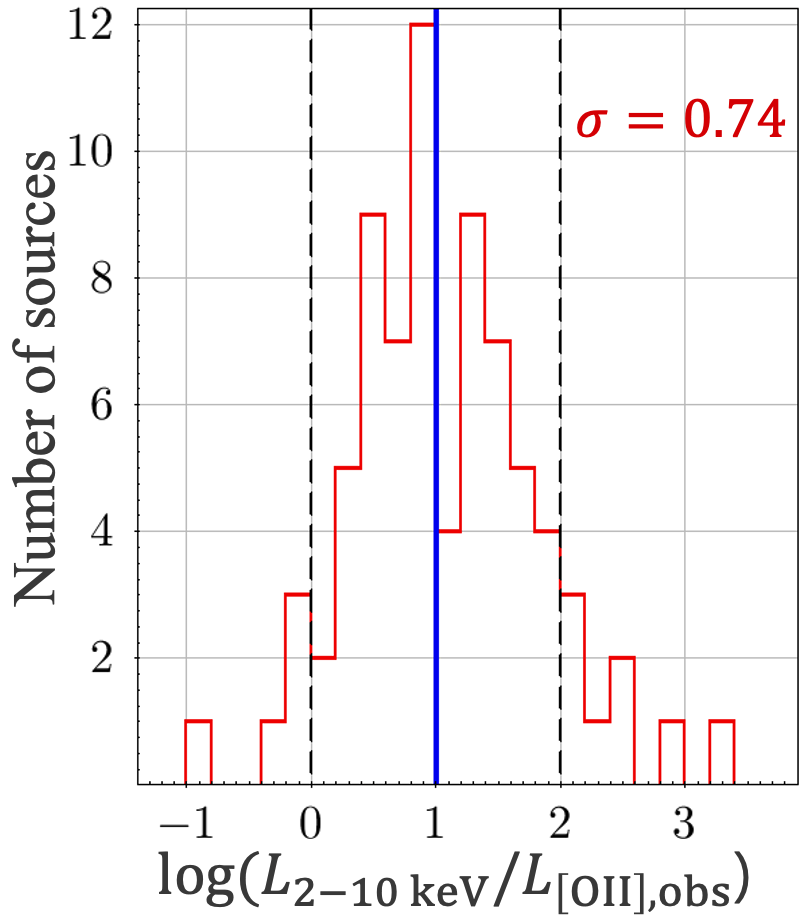}  
  \label{fig:5d}
\end{subfigure}

\caption{The relation between observed [O\,\emissiontype{II}] luminosity and hard X-ray luminosity (2-10 keV) for the NL AGNs (blue dots) in Panel (a) and for the OD AGNs (red dots) in panel (c). The solid line shows the ratio of hard X-ray to [O\,\emissiontype{II}] luminosity equal to 10, while dashed line shows the ratio value of 1 (lower dashed line) and 100 (upper dashed line). Panel (b) and (d) show distribution of $L_\mathrm{2-10 keV}/L_\mathrm{[OII], obs}$ for the NL AGNs and OD AGNs, respectively.The blue solid lines and the dashed lines corresponds to those in Panel (a) and (c). The standard deviation ($\sigma$) of the distribution is shown in the upper left of each histogram.}
\label{fig:LOII}
\end{figure*}

\subsection{Comparison of [O\,\emissiontype{II}] luminosity and hard X-ray luminosity}%$L_{\text{[O\,\emissiontype{II}]}}$ vs $L_{2-10\text{KeV}}$}
\label{sec:XOratio}
The X-ray emission is the unmistakable signature of AGN activity. We use the high luminosity constraint of $L_{0.5-10 \text{keV}}$ to select AGN. It is enough to conclude that star-forming galaxies do not strongly contaminate our sample. SF galaxy is known to has typical X-ray luminosity $\sim 10^{39}$ erg s$^{-1}$ \citep{Mineo} ). Therefore, the X-ray emission can be the main representative of AGN activity. 

In the subsection \ref{sec:hiOII }, we already distinguished the sources to be NL AGN and OD AGN regarding the AGN excess in their [O\,\emissiontype{II}] emission line compared to the typical value of SF galaxies. The excess observed in the NL AGN is the unmistakable evidence that [O\,\emissiontype{II}] emission of NL AGN was also coming from NLR as an AGN contribution. Meanwhile, the [O\,\emissiontype{II}] of OD AGN sources is concluded to originate only from its host galaxy. Therefore, the origin of [O\,\emissiontype{II}] emission is different for both samples.

Some previous studies already show a good relation between the [O\,\emissiontype{III}]$\lambda$5007 line and X-ray emission of AGN (\cite{Zakamska_2003}; \cite{Zakamska_2004}; \cite{Netzer_2006}). Thereupon, \citet{Zakamska_2003} presented a perfect linear relationship between [O\,\emissiontype{II}]$\lambda$3727 line and the [O\,\emissiontype{III}]$\lambda$5007 line in their AGN sample. It leads to an expectation of a good relation of [O\,\emissiontype{II}]$\lambda$3727 and hard X-ray (2-10 keV) as shown in the [O\,\emissiontype{III}]$\lambda$5007 line. The expected relation is clearly shown in our works, joining with the previous study that showed in Fig. \ref{fig:LOII}. 

The blue dots of NL AGNs in Fig. \ref{fig:LOII} (a)show a good relation between X-ray luminosity and [O\,\emissiontype{II}] luminosity. They found to have small standard deviation of $\sigma_{NL}= 0.4$ (shown in histogram \ref{fig:LOII} (b)) to the solid line which represents the ratio of X-ray luminosity and [O\,\emissiontype{II}] luminosity equal to ten ($L_{2-10 \text{keV}}=10L_{\text{[O\,\emissiontype{II}]}})$). It is in good agreement with a previous study by \citet{Netzer_2006} in the black crosses, and \citet{Tanaka_2012} in the green diamonds. The sample that was used by \citet{Tanaka_2012} are drawn from SDSS and classified as AGN using a BPT diagram. Their sample concentrates in a more local universe than our sample. They are lower luminous than ours. Meanwhile, \citet{Netzer_2006} which used higher redshift sample,  defined their sample as NL AGN based on the X-ray criteria in the luminosity and column density ($L_{2-10 \text{keV}} >10^{42}$ erg s$^{-1}$ and $N_\text{H}>10^{22} \text{cm}^{-2}$). Only two sources show an upper limit in X-ray luminosity shown with the blue down arrows. However, still these sources show the typical value of $L_{2-10 \text{keV}}/L_{\text{[O\,\emissiontype{II}]}}=10$. 
%Surprisingly, only a source that fall outside the ratio range of $L_{2-10 \text{keV}}/L_{\text{[O\,\emissiontype{II}]}})=1$ and $L_{2-10 \text{keV}}/L_{\text{[O\,\emissiontype{II}]}}=100$.

Meanwhile, the OD AGNs, shown in the red circles in the Fig. \ref{fig:5c}, distribute more scattered than NL AGNs with larger standard deviation of $\sigma_{OD}= 0.74$. This weak correlation among the OD AGNs gives us a proof that the majority of [O\,\emissiontype{II}] emissions were not coming from AGN. Instead, it is more likely to come from the star-forming regions of the host galaxies.

However, the OD AGN sample is still found to have a value consistent with the typical NL AGNs, which means that their host galaxy is basically bright enough to dilute their AGN optical signature. In opposite, the OD AGNs that placed above the black dashed line in Fig. \ref{fig:LOII} ($L_{2-10 \text{keV}}/L_{\text{[O\,\emissiontype{II}]}}>100$) are considered to have a nucleus (represent by X-ray luminosity) that is brighter than the nuclei in the typical NL AGN. The dilution scenario is difficult to explain since their AGN's light should be outshined the host galaxies' light. It leads us to find that a physical effect might depress their optical emission while remaining bright in X-ray.

\subsection{Host Galaxy properties}
\label{sec:hostgalprop}
We use secure morphological information of axial ratios and half-light radii in F814W of HST ACS for 47 NL AGNs and 177 OD AGNs obtained from the ZEST catalog \citep{Scarlata_2007}. For very compact objects, the instrumental point-spread function (PSF) could affect the axial ratio calculation. The limiting size/magnitude in which the effects of the PSF become important could make the preference of the morphological galaxy tends to be elliptical type. Despite, \citet{Scarlata_2007} stated that the ZEST catalog is able to measure the morphological parameters of the COSMOS sources without being affected by PSF convolution unless for sources with half-light radii of $r_{1/2}<0.17^{\prime\prime}$ (arcsecond). The number of sources in our sample with $r_{1/2}<0.17^{\prime\prime}$ are only four NL AGNs and four OD AGNs, which means that PSF unlikely to affect our result.

\begin{figure}
	\includegraphics[width=\columnwidth]{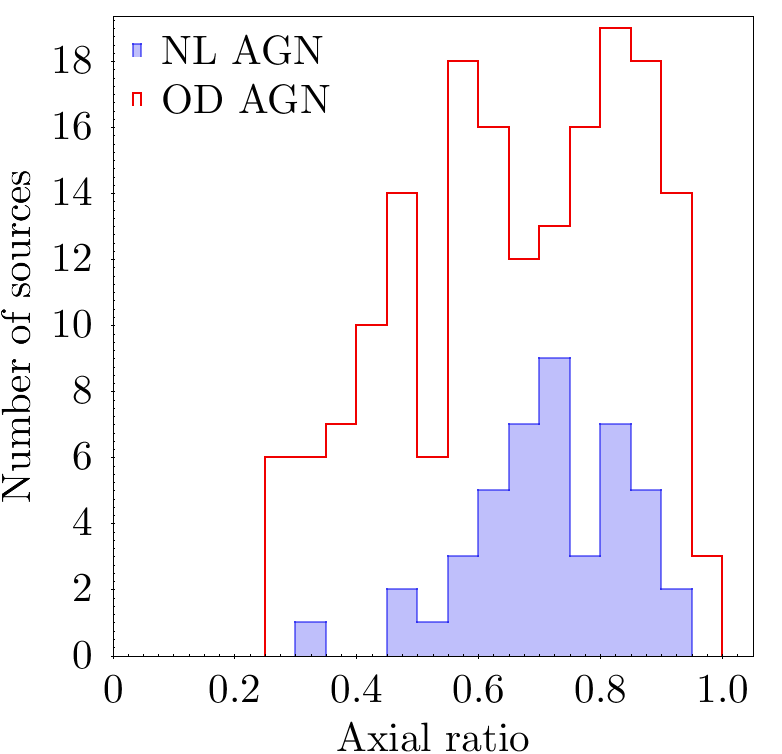}
    \caption{Axial ratio ($b/a$) histogram for the OD AGN (red line) NL AGN (shaded blue) sample. The axial ratio of OD AGNs and NL AGNs in our sample show similar range with no preference for specific value of inclination angle.}
    \label{fig:axisrat}
\end{figure}

We present the distribution of axial ratio for OD AGNs (red line histogram) along with NL AGNs (blue shaded histogram) in Fig. \ref{fig:axisrat}. \citet{Rigby_2006} shows that X-rays select AGNs in host galaxies with a wide range of axial ratios, but only AGNs that show optical emission (optically active) are observed in the most face-on host galaxies. It supports the idea that extranuclear dust in the host galaxy plays an important role in hiding the emission lines of OD AGNs. They compare the morphologies of 22 OD AGNs and nine optically active AGNs that contain BL AGN and NL AGN at $0.5<z<0.8$. Their OD AGNs distribute in the axial ratio range of $0.26 \le b/a \le 0.89$ while all their optically active AGNs have $b/a > 0.79$. In contrast with our sample, the OD AGNs and NL AGN have nearly similar ranges of axis ratio $b/a$, with the OD AGN mean value  $b/a=0.66\pm0.19$ and the NL AGN mean $b/a=0.7 1\pm 0.13$. The mean value is slightly different from the sample of \citet{Trump2009a} (OD AGN  $b/a=0.56\pm0.2$, NL AGN  $b/a=0.56\pm0.18$), which used a lower redshift sample ($z\lesssim1.0$).The NL AGNs in our sample, which are slightly higher redshift, most possibly show the face-on preferences. However, it is not shown in our NL AGNs or even in \citet{Trump2009a} that \citet{Rigby_2006} claims for their 'optically active' AGN sample. We agree with \citet{Trump2009a} that the difference comes from their sample definitions containing type 1 BL AGNs. In contrast, our sample only contains the NL AGNs. The morphological parameters that fit the BL AGN host galaxy could suffer from systematical errors due to bright nuclei. In particular, a BL AGN host could have an incorrectly high $b/a$ value. In the end, we conclude that the edge-on host is not the leading cause of the optical dullness among our OD AGN sample.

\begin{figure}
	\includegraphics[width=\columnwidth]{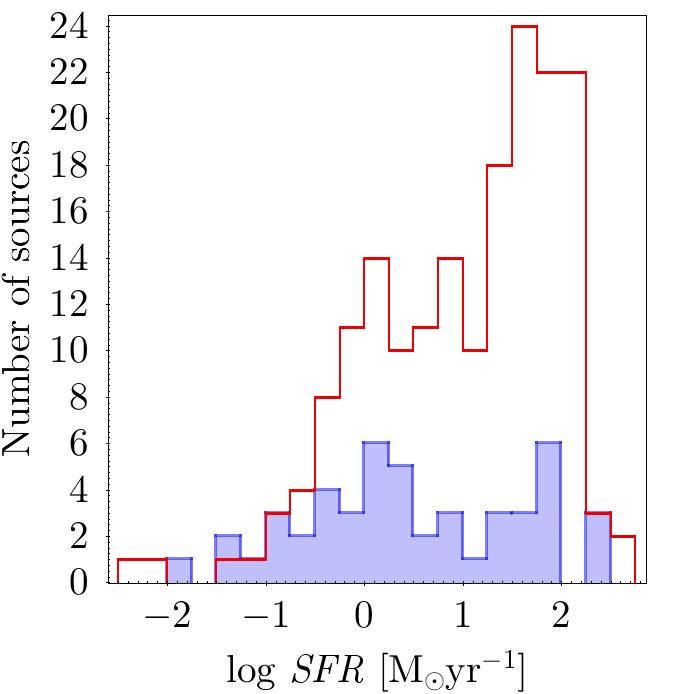}
    \caption{$SFR$ histogram of for the OD AGN (red line) NL AGN (shaded blue) sample. OD AGNs distributes in more larger $SFR$ galaxies. }
    \label{fig:SFR}
\end{figure}

For additional comparison of host properties, we present the star formation rates ($SFRs$) parameter of NL AGNs and OD AGNs in Fig. \ref{fig:SFR}. $SFRs$ were also adopted from COMOS2015 catalog as well as stellar masses that derived from the 30-band SED fitting by \cite{Laigle2016}. As written in Table \ref{table:AGNterm}, NL AGNs and OD AGNs have slightly different mean values ($\log M_*=10.78\pm 0.3 $ for OD AGNs, $\log M_*=10.82\pm 0.32 $ for NL AGNs). In other words, both samples distribute in the same stellar mass range. Therefore, we can use the $SFR$ parameter as an indicator of a galaxy actively forming stars (i.e., SF galaxies) since there is no significant trend toward mass in our samples. It is shown clearly in Fig. \ref{fig:SFR} that OD AGN (red line histogram) tends to be hosted by SF galaxies with typical $SFR$ larger than 10 M$_\odot$ per year. There are 101/180 (56\%) OD AGNs found to have $SFR>$ 10 M$_\odot$ yr$^{-1}$. Meanwhile, NL AGNs distribute more evenly and only have 16/48 (33\%) sources hosted by SFG. The mean value of $SFR$ for both sample also show significant different, which are $\log SFR=1.01\pm 0.96 $ for OD AGNs, $\log SFR=0.48\pm 1.1 $ for NL AGNs. The fact that OD AGNs have a tendency to be hosted by SF galaxies can be understood since we select the majority of this sample due to typical [O,\emissiontype{II}] emission of SF galaxies ( see our [O,\emissiontype{II}] excess classification in subsection \ref{sec:hiOII } for detail). Moreover, it is difficult to use only SFR parameter to explain the dilution scenario. We should compare the host galaxy and AGN properties to understand how the host galaxy dilutes the nuclei.

\section[Nature of OD AGN]{Nature of OD AGNs}
\label{sec:natureODAGN}
In this section, we will explore possible reasons for the optical dullness of our OD AGNs. Since we already examined host obscuration in section \ref{sec:hostgalprop}, we could exclude it from this discussion. Here, we present three principal natures for this dullness of OD AGNs, with some critical differences concerning the scenarios presented in previous studies:

\begin{enumerate}
    \item \textit{X-ray Obscuration}. Intrinsic X-ray emissions heavily attenuated by dust, rendering these X-ray AGNs unclassifiable by any optical methods in Section \ref{sec:Opticalclass} (\cite{Comastri_2002}; \cite{Civano_2007}).  %
    \item \textit{Star formation dilution}. Optical spectra taken by slit spectroscopy are contaminated by the emission from star-forming regions that changes the resulting emission line observed otherwise as a star-forming or normal galaxy (\cite{Moran_2002}; \cite{Pons2016}; \cite{Agostino2019}).%
    \item \textit{Low-accretion rate AGN.} The AGN is suspected of hosting a radiatively inefficient accretion flow (RIAF) which cannot adequately heat the NLR 
    (\cite{Yuan_2004}; \cite{Trump_2011b}).%
\end{enumerate}

\subsection{X-ray Obscuration}
\label{sec:Xrayobs}
One common explanation of optical dullness in OD AGN is obscuration by material near the central engine. \citet{Comastri_2002} suggested that OD AGN have precisely a similar physical engine as BL AGN or NL AGN, but with additional gas and dust obscuration covering a few parsecs from the nuclear source. This scenario could provide the necessary obscuration for blocking the ionizing radiation to excite in the NLR. However, obscuring the optical emission while remaining the X-ray emission bright would require extreme gas to dust ratios. It is because most OD AGNs were observed relatively unabsorbed with the hydrogen column density of $N_\text{H}<10^{22}$ cm$^{-2}$ (\cite{Severgnini2003}). 

To consider a necessary obscuration of the central engine, \citet{Civano_2007} proposed the obscuration caused by spherical Compton-thick gas cloud covering $\sim 4 \pi$ of the central region. This type of cloud's geometry will prevent the ionizing photons from escaping from the nuclear source and producing narrow line emission in the NLR. Therefore, we used the X-ray hardness ratio ($HR$) to estimate obscuration.

\begin{figure}
	\includegraphics[width=\columnwidth]{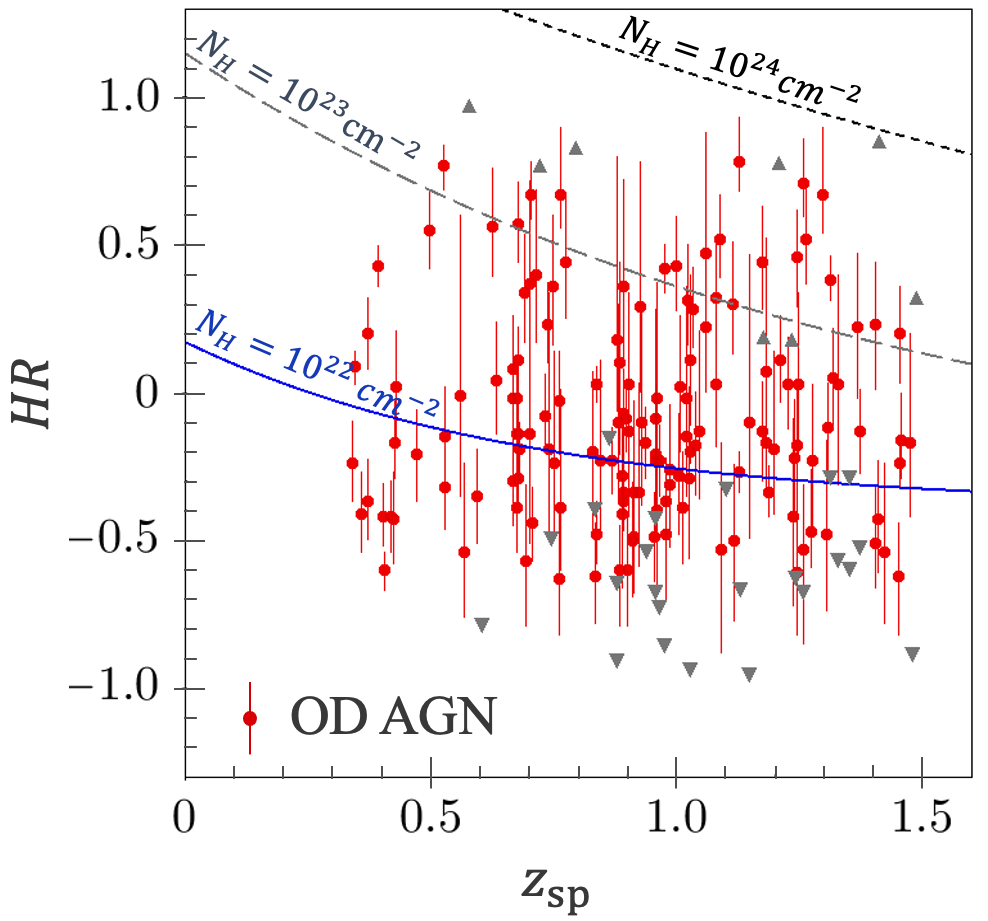}
    \caption{Hardness ratio (HR) relation with sectroscopic redshift ($z_\text{sp}$) for OD AGN sample in red dots symbol. The upper limit is shown in upside triangle while the lower limit is shown in downside triangle. Three curves of different column density ($N_\text{H}$) are plotted as comparison, obtained assuming a power-law spectrum with $\Gamma=1.8$: ($10^{22}$ $\text{cm}^{-2}$ in solid blue line, $10^{23} $ $\text{cm}^{-2}$ in grey long dashed line, $10^{24}$ $\text{cm}^{-2}$ in black short dashed line)}
    \label{fig:HR}
\end{figure}

The $HR$ statistically indicates the amount of absorption by assuming the primary power-law domination and the similar spectral index in all sources. The X-ray spectrum of an AGN with a higher intrinsic absorption level is much harder than an unabsorbed one because the absorber deferentially attenuates the soft X-ray emission. The HR value from the C-COSMOS legacy catalog was defined as

\begin{equation}
    HR=\frac{H-S}{H+S},
\end{equation}

where $H$ is the net count in the X-ray energy band of $2-10$ $\text{keV}$ and $S$ is obtained in the X-ray energy band of $0.5-2$ $\text{keV}$. The relation between hardness ratio and the hydrogen column density $N_\text{H}$ was also calculated in the C-COSMOS legacy catalog at all redshift. Fig. \ref{fig:HR} is shown the relation of $HR$ and $N_\text{H}$ along with the redshift. There are three curves of different column density ($N_\text{H}= 10^{22}$ $\text{cm}^{-2},$ $10^{23}$ $ \text{cm}^{-2},$ and $10^{24}$ $\text{cm}^{-2}$ from bottom to top respectively), obtained assuming a power-law spectrum with a photon index of $\Gamma=1.8$. As shown in \ref{fig:HR}, there are no sources laid above the curve $N_\text{H} = 10^{24}$ $\text{cm}^{-2}$ which is the lower limit of the Compton thick cloud. It gives us evidence that the Compton thick cloud hypothesis could not explain our sample's optical dullness. Moreover, the X-ray emission will be significantly absorbed and scattered again in higher energy above 10 \text{keV} (outside the Chandra X-ray telescope observable energy range). It means that high obscuration by Compton clouds is not possible in our sample. Our sample gives a bright X-ray emission lower than the 10 keV energy band.

\citet{Pons2016}, who also used samples in the COSMOS field, tested the presence of Compton thick obscuration using a different method. They looked for the hard X-ray and [O\,\emissiontype{III}] luminosity ratio ($L_{HX}/L_{\text{[O\,\emissiontype{III}]}}$) that are smaller than 0.2 among their sample sources. If the source has  $L_{HX}/L_{\text{[O\,\emissiontype{III}]}} < 0.2$, it is an evidence that the Compton thick cloud exists. A high absorption level near nuclear will depress the X-ray luminosity, while NLR's [O\,\emissiontype{III}] emission should be unaffected. In the end, they found all their sources lie above the Compton thick threshold ranging from 0.5 to 50. Therefore, it could be concluded if the X-ray obscuration is not the nature of their optical dullness sample, which agrees with our result. 

Even though the high X-ray obscuration is not among our sample, the moderate absorption ($N_\text{H} = 10^{22} $ $\text{cm}^{-2}$ $- 10^{24}$ $\text{cm}^{-2}$) might be an additional nature of optical dullness to the diluted OD AGNs. As shown in Fig. \ref{fig:HR}, most sources located above the blue solid curve $N_\text{H} = 10^{22} $ $\text{cm}^{-2}$ which is the typical limit to classify sources as NL AGN (moderately absorbed type) by X-ray emission. Meanwhile, we could not neglect the significant sources under the solid blue curve, a typical BL AGN value (unabsorbed type). We use the curve to separate our OD AGN sample as moderately absorbed (sources with column density  of $N_\text{H}>10^{22}$ cm$^{-2}$) and unabsorbed (sources with column density of $N_\text{H}<10^{22}$ cm$^{-2}$) for the following analysis. By this definition, we found 103 sources as absorbed OD AGN and 77 sources as unabsorbed OD AGN (including 20 upper limit sources).

\subsection{Star formation dilution}
\label{sec:SFdilution}

The other way to explain the nature of OD AGN's optical dullness is the dilution by the host galaxy to the nuclear emission. Dilution could happen due to the spectroscopic aperture reason. \citet{Moran_2002} used a large spectroscopic aperture to demonstrate that not so few nearby NL AGNs were diluted by their host galaxies. This scenario would come if the entire galaxy and occasionally even a nearby galaxy's companion were included in the spectroscopic slit. 

By half-light radii in the ZEST morphological catalog, we compare the galaxy size with the aperture size of the slit. Both z-COSMOS \citep{Lilly_2007} and Deimos 10K \citet{Hasinger_2018} catalogs used in this study used the multi-slit spectrograph, that has a single slit size equal to $1^{\prime\prime} \times 10^{\prime\prime}$. Let us only consider the slit width and neglect the slit length. We can see that 145 (80\%) of our OD AGNs have smaller half-light radii than the spectroscopy aperture, which means that they were observed by including the entire galaxy. The remaining 35 (20\%) OD AGN sources have sizes larger than the slit width. Thus, it is understandable that many `optically normal' AGNs could appear `optically dull' when we observed the entire galaxy placed in the spectroscopic apertures. 

\citet{Moran_2002} found $60\%$ of their sample appear optically dull due to larger observational aperture than the galaxy size, and that fraction of \citet{Trump2009a} was about $70\%$. Compared to their result, our sample shows a slightly higher fraction. It is possibly caused by our sample being taken at a higher redshift up to $z \sim 1.5$. Meanwhile, others are in the more local universe; \citet{Moran_2002} use sample up to $z\lesssim0.5$, and \citet{Trump2009a} used sample up to $z\lesssim1.0$. More distant galaxies will likely be captured within the entire galaxy and diluted more likely.

Other than H II regions in the host galaxy, there is another source of low ionized emission lines like [O\,\emissiontype{II}] known as diffuse ionized gas \citep{belfiore2016}. This gas is observed surrounding the plane of our galaxy and some other spiral galaxies. Not only H II regions but such diffuse ionized gas might dilute AGN emission lines. However, both sources were located in the extranuclear and we did not further separate the original dilution region in the host galaxy.

As discussed in section \ref{sec:XOratio}, we found that the [O\,\emissiontype{II}] luminosity from NL AGN gives a good relation with hard X-ray. At the same time, OD AGN show more scatters in Fig. \ref{fig:hiOII}. It is shown that [O\,\emissiontype{II}] luminosity among our OD AGNs is likely coming from the host galaxy without any AGN contamination. 

We introduce a parameter $R_\text{HX-[O\,\emissiontype{II}]}$ as the ratio of $L_{2-10\text{keV}}$ and $L_\text{[O\,\emissiontype{II}] }$ as written below,

\begin{equation}
    R_\text{HX-[O\,\emissiontype{II}]}= \log \frac{L_{2-10\text{keV}}}{L_\text{[O\,\emissiontype{II}]}}.
\end{equation}

This ratio estimates the AGN strength over the host galaxy among our OD AGN sample. Meanwhile, $R_\text{HX-[O\,\emissiontype{II}]}$ of NL AGN gives different meaning since there is an AGN contamination over the  [O\,\emissiontype{II}] emission line. It causes the [O\,\emissiontype{II}] luminosity of NL AGN is not appropriate to be the host galaxy's representative. However, the typical  $R_\text{HX-[O\,\emissiontype{II}]}$ value of NL AGNs by this work as well as two previous studies could give us an estimation of AGN strength. Consequently, we could find the bright host galaxy that possibly dilutes the OD AGN's light if we compare $R_\text{HX-[O\,\emissiontype{II}]}$ of OD AGN with that of NL AGNs ($R_\text{HX-[O\,\emissiontype{II}]}<2$).

\begin{figure}
	\includegraphics[width=\columnwidth]{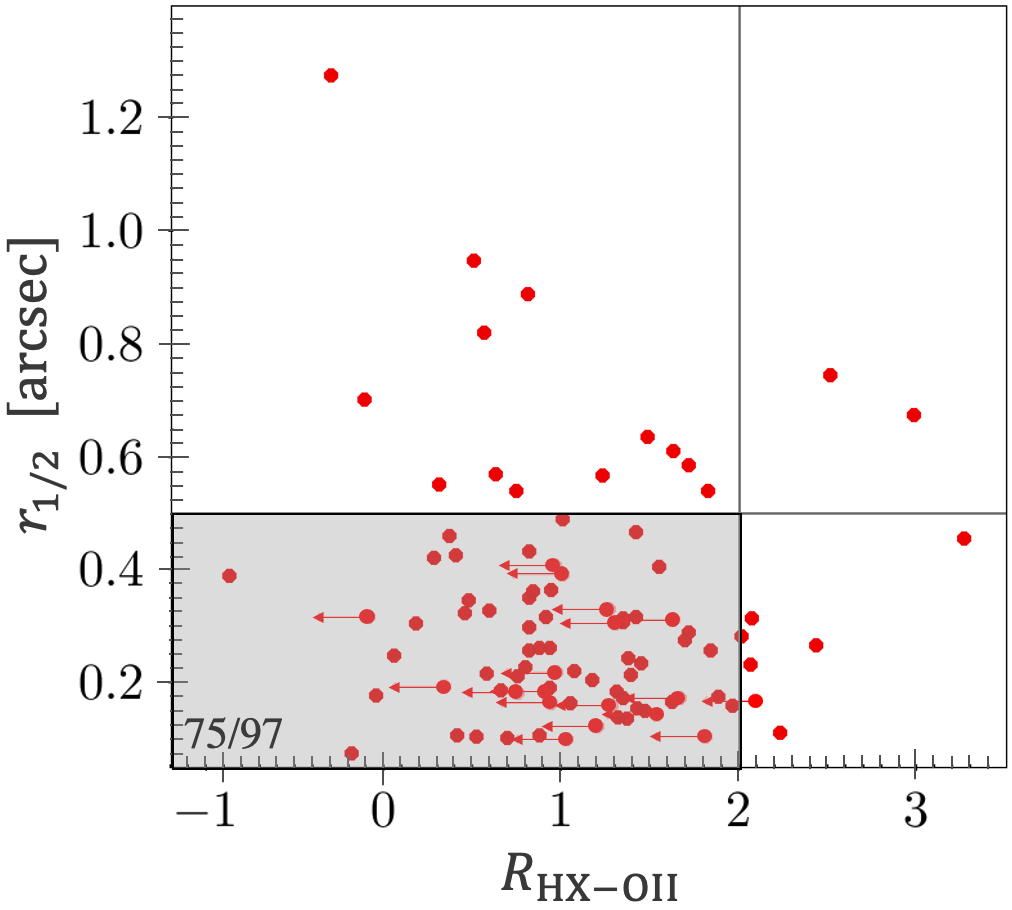}
    \caption{The $R_\text{HX-[O\,\emissiontype{II}]}$ as AGN domination over host galaxy vs. the half-light radii ($r_{1/2}$). The horizontal line represents the size of the spectroscopic aperture (slit size $\sim 1^{\prime\prime}$) while the vertical line represents the limit of the typical value of 
    NL AGN's ($R_\text{HX-[O\,\emissiontype{II}]} =2$).}
    \label{fig:dilution}
\end{figure}

Fig. \ref{fig:dilution} basically shows the dilution effects among our OD AGNs with comparing $R_\text{HX-[O\,\emissiontype{II}]}$ with half-light radii. The region under the horizontal line represents that the spectroscopic aperture is comparable to the host galaxy's size. We have found that the spectroscopic aperture dilutes 80\% of our OD AGN sample. However, bright AGN might be suffering from this dilution scenario. Hereabouts, we also consider the AGN light compared to the host galaxy light, represented by the X-ray and [O\,\emissiontype{II}] luminosity ratio as $R_\text{HX-[O\,\emissiontype{II}]}$. Hence, We put the constrain in the $x$-axis equal to 2 $ (L_{2-10\text{keV}}=100\times{L_\text{[O\,\emissiontype{II}]}})$ to state the AGN light that possibly overwhelms the galaxy light since it is brighter than the typical NL AGNs.

The OD AGNs, which has $R_\text{HX-[O\,\emissiontype{II}]}$ value more than the vertical line in Fig. \ref{fig:dilution}, should be brighter and could overshine the host galaxy. Thus, along with the additional effect of large observational aperture, we found 75 out of 97 ($\sim 77\%$) sources likely to be diluted by their host galaxies ( shown in grey shaded area). Moreover, the sources with a  $R_\text{HX-[O\,\emissiontype{II}]}$ value more than the vertical line ($R_\text{HX-[O\,\emissiontype{II}]} =2$) and have a larger size ($r_{1/2} >0.5 ^{\prime\prime}$) are unlikely to have a host galaxy's dilution to explain their optical dullness. It gives us an indication that there is another nature to their optical dullness. 

\subsection{Low-accretion rate AGN}
\label{sec:RIAFhypo}
AGNs with low accretion rates are expected to be optically underluminous, with very weak or missing emission lines \citep{Narayan_1998}. In this section, we will investigate the third reason for the optical dullness by estimating the specific accretion rate using the Eddington ratio parameter $\lambda_{\text{Edd}} \equiv L_\text{bol}/L_\text{Edd}$. Here, $L_\text{bol}$ will be estimated using the X-ray luminosity. In contrast, the Eddington luminosity is derived from the black hole mass.

Estimating BH masses for AGNs without broad emission lines requires secondary estimators. We use the well-studied correlation between BH and bulge mass of the host galaxy\citep{Marconi_2003}. 

\begin{equation}
    \log(M_\text{BH})=8.12+1.06(M_\text{bulge}-10.9),
\end{equation}

where $M_\text{BH}$ is the blackhole mass and $M_\text{bulge}$ is stellar mass of the bulge component in unit of $M_{\odot}$. More recent studies estimated the $M_\text{BH}$ using the K-band luminosity relation (\cite{Graham_2007}; \cite{Vika_2011}; \cite{Lasker_2013}) instead of the stellar mass. We also tried the K-band luminosity relation in our sample and it still gives similar range of $M_\text{BH}$. However, we prefer to keep our calculation using  $M_\text{BH}-M_\text{bulge}$ relation which is systematically tighter relation than that  the $M_\text{BH}-L_\text{K,bulge}$ relation, where $L_\text{K,bulge}$ is the $K$-band luminosity of the host galaxy's bulge. It is expected if $L_\text{bulge}$ correlates with $M_\text{BH}$ because of its dependence on $M_\text{bulge}$ through the stellar mass to light $M/L$ ratio. 

Furthermore, we combine the total galaxy's stellar mass from \citet{Laigle2016} with bulge fractions from morphological classification by ZEST catalog (\cite{Scarlata_2007}) in order to determine the $M_\text{bulge}$. By adopting \citet{Lusso_2012}, we decided a fraction of the bulge-to-total ($B/T$) based on the morphological type as well as the bulgeness level, which are presented in the ZEST catalog as follows : 

\begin{enumerate}
    \item elliptical :$B/T$=1,
    \item bulge-dominated disc: $B/T$=0.75,
    \item intermediate-bulge disc: $B/T$=0.5, and
    \item disc-dominated: $B/T$=0.25.
\end{enumerate}

\begin{figure*}
\begin{subfigure}{.5\textwidth}
  \centering
  \caption{}
  % include first image
  \includegraphics[width=\textwidth]{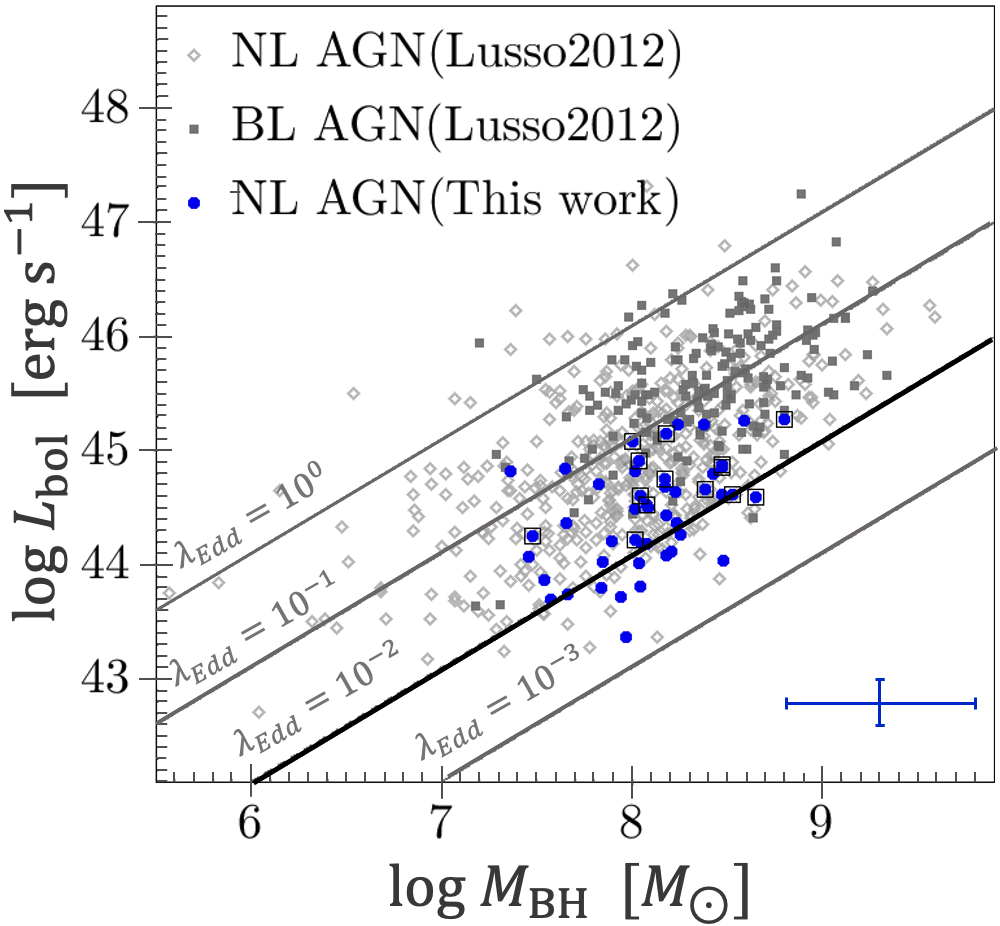}  
  \label{fig:10a}
\end{subfigure}
\begin{subfigure}{.35\textwidth}
  \centering
  \caption{}
  % include second image
  \includegraphics[height=1.27\textwidth]{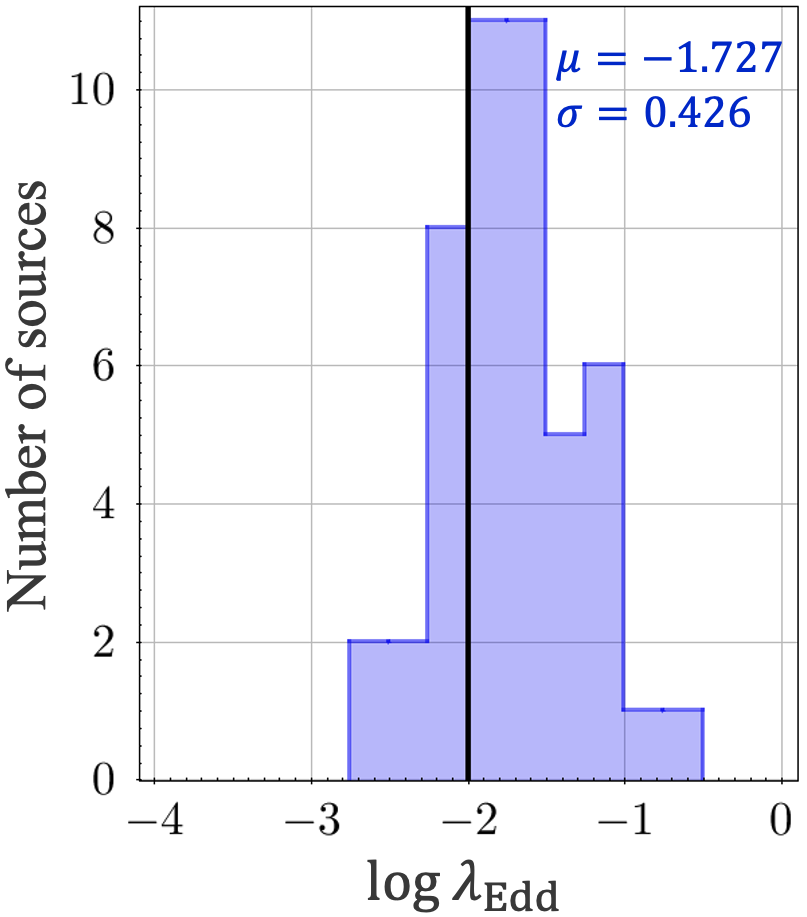}  
  \label{fig:10b}
\end{subfigure}

\begin{subfigure}{.5\textwidth}
  \centering
  \caption{}
  % include third image
  \includegraphics[width=\textwidth]{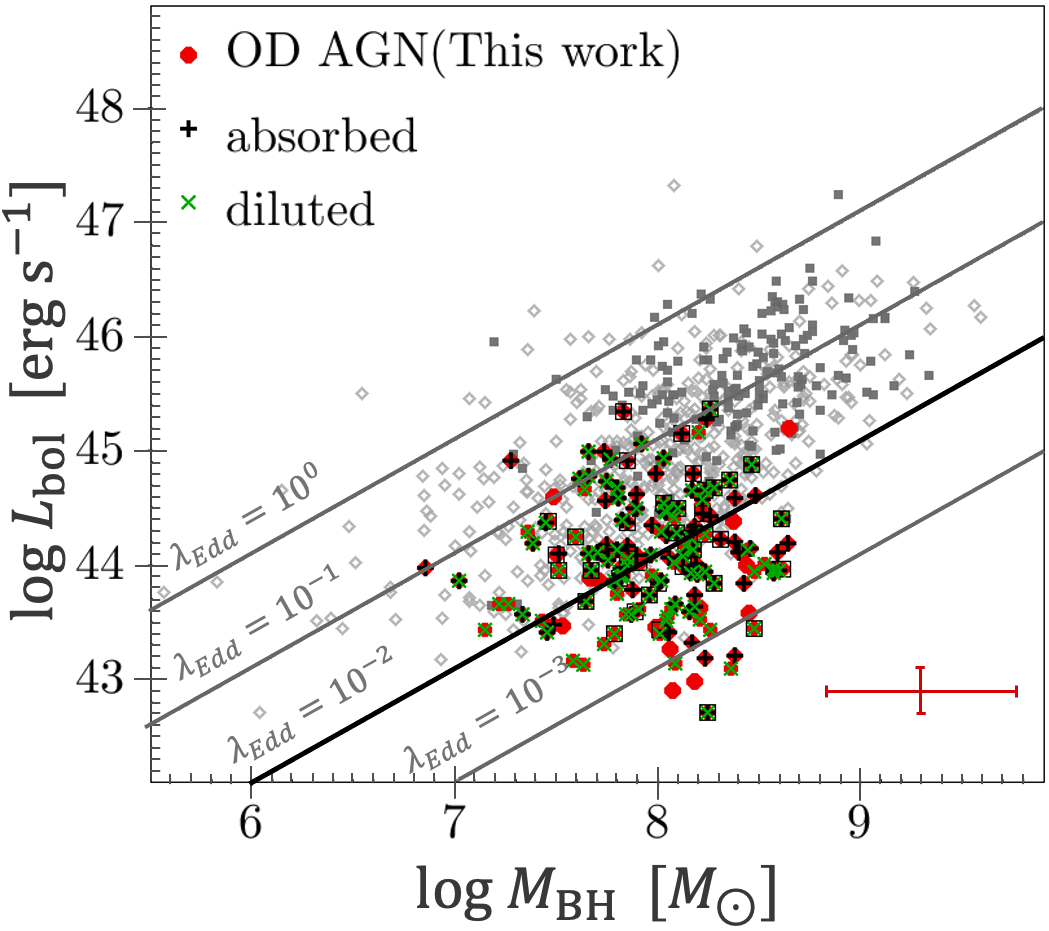}  
  \label{fig:10c}
\end{subfigure}
\begin{subfigure}{.35\textwidth}
  \centering
  \caption{}
  % include fourth image
  \includegraphics[height=1.25\textwidth]{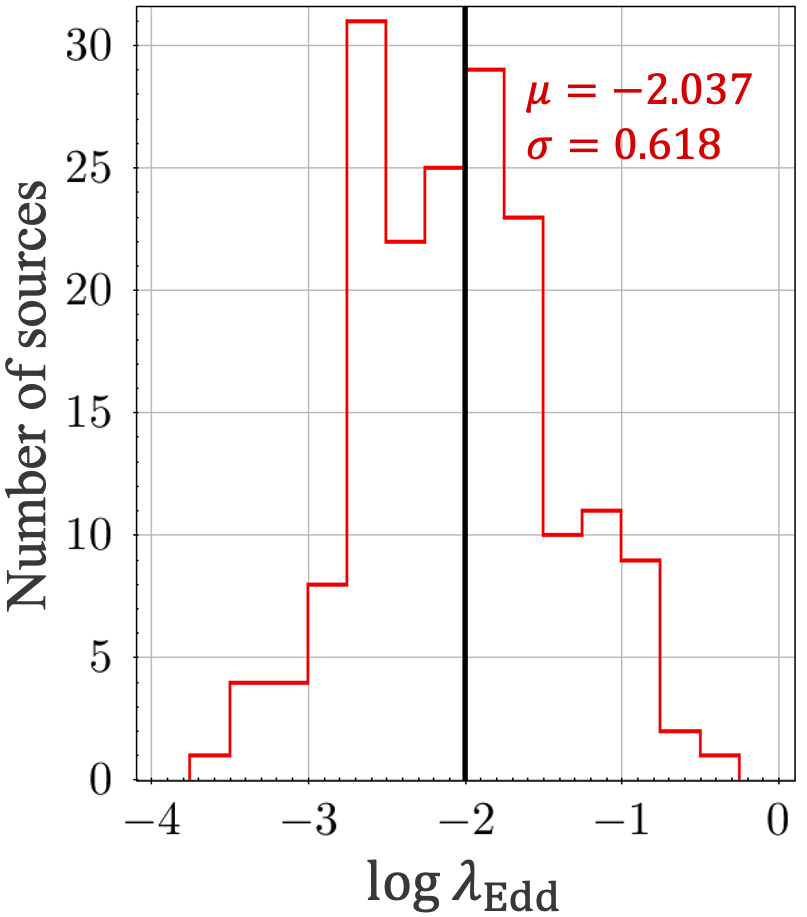} 
  \label{fig:10d}
\end{subfigure}

\caption{The diagram of BH mass ($M_\text{BH}$) vs bolometric luminosity ($L_\text{bol}$) for our sample compared with \citet{Lusso_2012} in grey diamonds and squares symbols. Panels (a) and (c) show the diagram of NL AGNs (blue dots) and OD AGNs (red dots), respectively. The typical error of the data shown in the error bar in lower right of the figure. The solid lines show loci of the same Eddington ratio of $\lambda_\mathrm{Edd}=10^{0}$, $10^{-1}$, $10^{-2}$, and $10^{-3}$ (from upper left to bottom right). Object with upper limit of $M_\mathrm{BH}$ are shown with black squares. The diluted OD AGNs and absorbed OD AGNs are shown in black pluses and green crosses, respectively. Panels (b) and (d) show the histogram of the Eddington ratio for NL AGNs and the OD AGNs, respectively. The mean value ($\mu$) and the standard deviation ($\sigma.$) of the distribution are shown in each panel. }
\label{fig:eddrat}
\end{figure*}

%The uncertainties in $B/T$ are typical $\sim 0.3$, with the limit that 0 $\le$  $B/T$ $\le$1. 
In the ZEST catalog, 142 OD AGNs and 34 NL AGNs are classified as one of the above types of morphology. Meanwhile, the remaining 38 OD AGNs and 14 NL AGNs are classified as irregular types or unable to determine. For these sources, we employed the $M_{*}$ as $M_\text{bulge}$ in the BH mass estimation ($B/T=1$) and considered them as upper limit value (the open square symbol in Fig. \ref{fig:eddrat}).

We estimated the bolometric luminosity of AGN from the X-ray luminosity by applying a bolometric correction. We have used hard X-ray luminosity ( $L_{2-10\text{keV}}$), while those with the upper limit of hard X-ray, the soft X-ray luminosity ($L_{0.5-2\text{keV}}$) was used. Assuming relations by \citet{Lusso_2012}, the bolometric corrections were derived as follows:
\begin{equation}
    \log \left( \frac{L_\text{bol}}{L_{2-10\text{keV}}}\right)=1.256+0.23\mathscr{L}+0.05 \mathscr{L}^2-0.001\mathscr{L}^3,
\end{equation}
and
\begin{equation}
    \log\left(\frac{L_\text{bol}}{L_{0.5-2\text{keV}}}\right)=1.399+0.217\mathscr{L}+0.009 \mathscr{L}^2-0.010\mathscr{L}^3,
\end{equation}

where $\mathscr{L}=\log L_\text{bol} -12$, and $L_\text{bol},$ $ L_{2-10\text{keV}},$ $L_{0.5-2\text{keV}}$ are in the unit of $L_{\odot}$. This bolometric correction are performed for NL AGN which assume the instrinsic nuclear luminosity are coming from the IR and X-ray luminosities. The mid-IR luminosity is considered as an indirect probe of the accretion disc optical/UV luminosity.

%Meanwhile, the Eddington luminosity is directly related to the BH mass as 
%\begin{equation}
%    L_\text{Edd}=1.25 \times 10^{38} \left(\frac{M_\text{BH}}{M_{\odot}}\text{erg s}^{-1}\right),
%\end{equation}

In Fig. \ref{fig:eddrat}, the bolometric luminosities ($L_\text{bol}$) are plotted as a function of BH masses ($M_\text{BH}$) for OD AGN ((a)) as well as the NL AGN ((b)) in this work. As a comparison, we also put the result from the previous study by \citet{Lusso_2012} in a grey symbol (BL AGN as filled square while NL AGN as open diamond). We add the upper limit value of black hole mass (the sources with irregular or no morphological type in the ZEST catalog) in an open square symbol in the panel (a) and (c) of Fig. \ref{fig:eddrat}. The upper limit values which are showed in open squares distributes in the same range as the remaining sources. The diagonal lines represent the relation between $L_{\text{bol}}$ and $M_{\text{BH}}$ at different Eddington ratios ($\lambda_{\text{Edd}}=10^0,10^{-1},10^{-2},$ and $10^{-3}$). Typical errors of $M_{\text{BH}}$ and $L_\text{bol}$ are measured as 0.46 and 0.2, respectively. 

As we can see in Fig. \ref{fig:10a}, our NL AGNs ($\log$ $\lambda_\text{Edd}=-1.72\pm 0.43 $) show similar distribution as the NL AGNs from \citet{Lusso_2012} ($\log$ $\lambda_\text{Edd}=-1.72\pm 0.59$) of which has slightly lower redshift sample ($z<1.2$). The BL AGN sample is shown in a similar range of Eddington ratio but higher in black hole mass as well as the bolometric luminosity. Meanwhile, the OD AGNs in Fig. \ref{fig:10c}, show more scatter distribution than NL AGNs. The Eddington ratio of OD AGNs ($\log$ $\lambda_\text{Edd}=-2.04\pm 0.62$) are shown slightly lower than NL AGN. However, we do not find any difference between the diluted and undiluted populations among OD AGN distribution. The Eddington ratio of undiluted sample ($\log$ $\lambda_\text{Edd}=-1.96\pm 0.68$ ) show slightly larger than diluted sample ($\log$ $\lambda_\text{Edd}=-2.00\pm 0.6$).

\citet{Trump_2011b} mentioned that the accretion rates are supposed to be $\lambda_\text{Edd}\lesssim 10^{-2}$ to avoid BLR forming due to RIAF disk in the inner radii of the accretion disk. As result, \citet{Trump_2011} and \citet{Pons2016} showed all of their OD AGNs had lower accretion rates than the limit of $\lambda_\text{Edd}\lesssim 10^{-2}$. Instead, our result gives more scattered value in accretion rates than those studies. In our OD AGNs, 85 sources show accretion rates larger than the limiting Eddington ratio of $10^{-2}$ showing the typical accretion rate of NL AGNs. Meanwhile, 95 OD AGNs still show low accretion rates as typical RIAF disk sources. 

In total, OD AGNs show systematically higher accretion rates than previous studies. We consider this difference to come from the sample selection process. \citet{Trump_2011} had spectroscopic data with observational limit of $i_\text{AB} \lesssim 23$ mag which might lack of fainter galaxies population compare to ours that give observational limit of $i_\text{AB} \lesssim 25$ mag.  We also selected more distant galaxies in a deeper observation which is most likely to select lower mass galaxies. Furthermore, in contrast with our sample selection, their sample eliminated the emission-line galaxies from the OD AGN sample., probably due to SF activity. As a result, our OD AGNs tend to have smaller $M_{\text{BH}}$ galaxies and therefore show higher accretion rates. The difference with \citet{Pons2016} is most probably due to the X-ray luminosity limit to select the X-ray AGN being slightly lower than ours. They set the minimum values of $L_X \sim 10^{42}$ erg s$^{-1}$ which is giving a lower bolometric luminosity then resulting in lower accretion rates.

Nevertheless, we could find a more significant number of OD AGNs that show a lower accretion rate of $\lambda_\text{Edd}\lesssim 10^{-2}$ than previous studies. Even though the diluted sources may be a majority of these lower luminosity OD AGNs, we could still obtain a considerable OD AGN sample as a good candidate of Radiatively Inefficient Accretion Flow (RIAF) sources. The RIAF sources here are defined as the truncated disk in the innermost region of the optically thick disk. Thus, the optical and UV emission is very weak due to the absence of an optically thick disk at small radii, 

Some OD AGNs in our sample have low hydrogen column densities of $N_\text{H}< 10^{22}$ cm$^{-2}$, which is the typical value of unabsorbed BL AGNs. \citet{Tran_2003} discovered similar sources called ``naked NL AGNs'', which were characterized by a lack of a broad emission line even in a spectropolarimetry observation. These sources cannot be explained by the simple unified model in viewing angle. They may have fundamentally different accretion physics like RIAF.

Our OD AGNs with low accretion rates can not be securely confirmed to host a RIAF based only on the accretion rate alone. We are required to discuss the sources most likely to have RIAF by comparing them with the RIAF model. RIAF model has characteristic emission in two primary wavelengths: radio and X-ray.

RIAF are common to be found in nearby low luminosity AGNs. The occurance of outflows in RIAFs can explain the weaker radiative outputs and large ratio of the X-ray to radio luminosities of low luminosity AGNs (\cite{Dimateo}). In the study of \citet{Nemmen_2014} presented a sample of low luminosity AGNs that have observed broadband spectrum are fitted into the RIAF model. Their sample had lower Eddington ratio than ours with typical value of $\lambda_\text{Edd}\sim 10^{-5}$ due to their X-ray low luminosity ($L_X\sim 10^{40}$ erg s$^{-1}$). Similar to our OD AGN sample, these low luminosity AGNs do not seem to have the optical signature of standard thin disk AGNs. 

An approach to confirm RIAF by broadband spectra SED fitting is likely to work on our sample. There will be nine sources with a lower accretion rate and do not suffer from dilution or high absorption that will be an appropriate target to be fitted. Moreover, suppose these sources are proved to host a RIAF disk. In that case, it leads us to conclude that OD AGN is a more luminous and distant version of low luminosity AGN. It supports the scenario regarding the evolution of SMBH to be more active in the more distant universe. Furthermore, it can support the idea of changing accretion disk flow from standard disk in higher redshift into RIAF truncated disk in lower redshift.

RIAFs have also been demonstrated to naturally accommodate hard X-ray emission, along with gamma-ray emission. The gamma-ray emission in the innermost of our Galaxy (the confirmed RIAF source) is considered to be consistent with inverse Compton scattering by the RIAF disk. Low luminosity AGNs are also detected by the Fermi Large Area Telescope as potential gamma-ray sources (\cite{abdoa}). Using the RIAF model, we can predict the gamma-ray spectrum of our sources. It will be a good comparison with future Fermi detections. 

% ***************************************************
% Conclusion
% ***************************************************
\section{Summary \& Conclusion}
\label{sec:summary}
We have presented optical diagnostic tools based on emission line and galaxy properties, allowing us to find the AGN signature in our sample up to redshift 1.5. Using the multiwavelength catalog by the COSMOS project, we were able to study the NL AGN and OD AGN properties. We found 180 sources without AGN signatures in optical wavelengths, corresponding to a fraction of 60$\%$ of our X-ray-selected AGN sample.  

One diagnostic tool for finding the AGN signature is finding the excess in the [O\,\emissiontype{II}] emission line. Unsurprisingly, we found a good relation of the [O\,\emissiontype{II}] luminosity ($L_\text{[O\,\emissiontype{II}]}$) and the X-ray luminosity ($L_{2-10\text{KeV}}$) among NL AGNs. Along with the previous study by \citet{Tanaka_2012} and \citet{Zakamska_2003} data, the typical value X-ray to [O\,\emissiontype{II}] ratio ranges of $0<R_{HX,[O\,\emissiontype{II}]}<2$. It shows us that the [O\,\emissiontype{II}] emission in these NL AGNs was coming from not only the host galaxy but also coming from the NLR of AGN. As one might expect, its is tricky to use [O\,\emissiontype{II}] emission to derive host galaxy properties in NL AGN since it is highly contaminated by AGN. Meanwhile, The relation cannot be addressed in our OD AGN sample. It is confirmed that the [O\,\emissiontype{II}] emission is only coming from host galaxies.

The dilution scenario would easily explain the nature of optical dullness. By comparing the $R_{HX,[O\,\emissiontype{II}]}$ value of our OD AGN sample and the typical NL AGN, we could present the OD AGNs that overwhelmed by the hosts. Besides that, the dilution can occur if we observe the entire galaxy placed in the spectroscopic aperture. The spectral observation of the entire galaxy from the bright host galaxy is possible to overshine the nuclei of OD AGNs. Among our OD AGN case, the host galaxy dilution can explain $\sim 70\% $ of optical dullness. This fraction is similar with other authors proposal (\cite{Moran_2002}; \cite{Trump2009a}).

Meanwhile, the high obscuration is least probably to explain the lack of optical AGN signatures in our case. The orientation, represented by the axial ratio, the OD AGN, and NL AGN, gives no difference in distribution ranges from 0.2 to 1 (edge on to face on view). Besides that, based on hardness ratio value, the X-ray obscuration is less dense than typical Compton-thick clouds for all the samples. However, the obscured phase might be addressed to the different nature of the OD AGN. Most of our undiluted sources are found to have absorbed X-rays. We can not rule out this fact.

The most interesting is whether optical dullness is due to the different intrinsic structures causing the low accretion rate. Hence, we estimated the Eddington ratio using black hole mass and bolometric luminosity estimation. The Eddington ratio is essential for detecting and identifying a galaxy as an AGN. We found that our estimators give an excellent result compared with previous work by \citet{Lusso_2012}, which have similar ranges and mean values of NL AGN in our work. Unlike \citet{Trump_2011} and \citet{Pons2016} that show all their OD AGNs give lower accretion rate than of $\log$ $\lambda_\text{Edd} \sim - 2$, we still got half of our OD AGNs distribute in a higher Eddington ratio similar to typical BL AGN and NL AGN ($\log$ $\lambda_\text{Edd} \ge -2$) while the remaining half are below the limit. According to the sample selection, we come across more distant and fainter galaxies, most likely to result differently.

Finally, we found 9 sources that can not be explained by any dilution and do not seem to be obscured as an appropriate candidate of RIAF galaxies. We consider these OD AGNs most probably possess RIAF disks. It is helpful to place OD AGNs in the context of other accreting black hole sources, which better estimate the AGN population.

% ********* Enter your text below this line: ********

\begin{ack}
We would like to thank the referee for his/her detail comment and suggestion, which helped to improve this paper. We also would like to thank Prof. Yoshiaki Taniguchi and Prof. Masayuki Akiyama for their sincerity reviewing our draft and giving useful feedback. Itsna would also like to acknowledge the support from Japanese Goverment (Ministry of Education, Culture, Sports, Science, and Technology or MEXT) scholarship for her studies.

This work is based on observation taken by COSMOS collaboration. More information on the COSMOS survey is available at http://www.astro.caltech.edu/~cosmos. It is a pleasure to acknowledge the available online archive for the COSMOS datasets. Some of the datas present in this work also comes from previous study by Dr. Masayuki Tanaka and Dr. Elisabetta Lusso. We would like to thank and are really appreciate their kindness for sharing their work data with us. This research also made use of Astropy (http://www.astropy.org), a community developed core Pyhton package for Astronomy and TOPCAT (http://www.starlink.ac.uk/topcat/).
\end{ack}

\bibliographystyle{aa}
\bibliography{./Bibliography}

\end{document}